\documentclass[preprint, 12pt]{elsarticle}
\usepackage{amsmath, physics, amssymb, amsthm}
\usepackage{algorithm}
\usepackage{algorithmic}
\usepackage{mathtools}
\usepackage{tikz}
\usetikzlibrary{mindmap}
\usetikzlibrary{shapes}
\usepackage[center]{caption}
\usepackage{epstopdf}
\usepackage{subcaption}
\usepackage{booktabs}
\usepackage{float}
\usepackage{pdfcomment}

\newtheorem{theorem}{Theorem}
\newtheorem{proposition}{Proposition}

\newproof{IEEEproof}{Proof}

\DeclareMathOperator*{\argmin}{arg\,min}

\graphicspath{{images/}}

\journal{Signal Processing}

\begin{document}
	
	\begin{frontmatter}
	\title{On Pooling-Based Track Fusion Strategies : Harmonic Mean Density }
	\author{Nikhil Sharma%
	}
	\ead{sharmn66@mcmaster.ca}
	\author{
		Ratnasingham ~Tharmarasa
	}
	\ead{thamas@mcmaster.ca}
	
	\author{
		Shovan Bhaumik
	}
	\ead{shovan.bhaumik@iitp.ac.in}
	
	\author{
		Thiagalingam Kirubarajan
	}
	\ead{kiruba@mcmaster.ca}
	
		\begin{abstract}	
				
				In a distributed sensor fusion architecture, using standard Kalman filter (naive fusion) can lead to degraded results as track correlations are ignored and conservative fusion strategies are employed as a sub-optimal alternative to the problem. Since, Gaussian mixtures provide a flexible means of modeling any density, therefore fusion strategies suitable for use with Gaussian mixtures are needed. While the generalized covariance intersection (CI) provides a means to fuse Gaussian mixtures, the procedure is cumbersome and requires evaluating a non-integer power of the mixture density. In this paper, we develop a pooling-based fusion strategy using the harmonic mean density (HMD) interpolation of local densities and show that the proposed method can handle both Gaussian and mixture densities without much changes to the framework. Mathematical properties of the proposed fusion strategy are studied and simulated on 2D and 3D maneuvering target tracking scenarios. The simulations suggest that the proposed HMD fusion performs better than other conservative strategies in terms of root-mean-squared error while being consistent. 
		\end{abstract}

		\begin{graphicalabstract}
			\begin{figure}
				\centering
				\begin{tikzpicture}[grow cyclic, text width=2.7cm, align=flush center,thick,scale=0.75, every node/.style={scale=0.5},
					level 1/.style={level distance=3.5cm,sibling angle=72},
					level 2/.style={level distance=2.4cm,sibling angle=60}
					]
					\node{\textbf{Track to Track Fusion}}
					child{node{Exact}
						child{node{Distributed Track Filtering \cite{govaers2012exact} }}
						child{node{Channel Filter \cite{chang2010analytical}}}
						child{node{Globalized Likelihood \cite{govaers2011globalized}}}
					}
					child{node{Decorrelation}
						child{node{Tracklets \cite{drummond1997tracklets}}}	
						child{node{Quasi-Tracklets \cite{drummond2002track}}}
						child{node{IMM Decorrelation \cite{acar2020decorrelation}}}
					}
					child{node{Pooling Strategies \cite{ajgl2014conservativeness}}
						child{node{Geometric Averaging \cite{mahler2000optimal}}
						child{node{Covariance Intersection \cite{reinhardt2012closed}}}	
						}
						child{node{Arithmetic Averaging \cite{li2020arithmetic}}}
						child{node{\underline{Harmonic Averaging}}}
					}
					child{node{Information form}
						child{node{Generalized IMF \cite{tian2010algorithms}}}
					}
					child{node{Sampling based}
						child{node{Ensemble KF \cite{sirichai2013using}}}	
					}
					
					;
				\end{tikzpicture}
			\end{figure}
		\end{graphicalabstract}
		
	\end{frontmatter}

	\section{Introduction}
	A distributed data fusion scenario comprises several sensor platforms connected to a node where the local information from sensors is fused. Since a sensor can capture multiple detections in the presence of clutter, sending raw measurement is avoided. Instead, the sensor communicates estimated target states (tracks), which is why the term track-to-track fusion (T2TF) is commonly adopted.
	
	Tracks estimating the same target trajectory in T2TF are never independent, even though the local measurement errors can be assumed uncorrelated \cite{bar1981track}. The dependence emerges due to the common process noise, and temporal correlation between tracks arriving from the same platform. Ignoring this correlation can lead to a degraded performance as the resulting estimate is optimistic \cite{bar1995multitarget}. The optimism is due to the aggregation of \textit{``common information"} in sensor estimates. This common information has to be accounted for and removed from the fused information to avoid track divergence.
	
	The approaches used in accounting for cross-correlation for the last three decades can be categorized into:
	\begin{enumerate}[1.]
		\item  Optimal solution based on evaluating exact recursion for correlation, which can later be accounted for in fusion. This strategy would require extra information from local sensors like Kalman gain, sensor, and motion model parameters \cite{bar1981track} \cite{chang2008scalable}. 
		\item Conditionally optimal methods like information matrix fusion, which would also require knowledge of previous local estimates \cite{chang2002performance}. Such methods are optimal only for a full communication rate, i.e. the fusion occurs every time the local estimates are updated.  
		\item  Tracklet and quasi-tracklet fusion methods are based on track decorrelation by marginalizing the joint density. Here again, knowledge about a track's previous estimate is needed.
		\item Pooling based solutions like the generalized covariance intersection, which utilizes a geometric mean (GM) or the arithmetic mean (AM) interpolation of individual track densities \cite{hurley2002information,li2020arithmetic}. 
	\end{enumerate}
	Generalized covariance intersection/Chernoff fusion/geometric averaging has attracted researchers due to its robustness and stability \cite{battistelli2014kullback}. An essential aspect of this method is that it essentially constitutes a density function which is a log-linear combination of the individual local posterior densities \cite{hurley2002information}, such that the propagation of the ``common information" is avoided \cite{bailey2012conservative}. Due to computational intensiveness of geometric mean for Gaussian mixtures, an alternative was proposed in \cite{li2020arithmetic, abbas2009kullback}.

	With the advent of computation and embedded technologies, the use of mixture densities has been adopted to deal with multi-modal, non-symmetric noise distributions and multiple-model approaches to filtering.  
	Moreover, the Gaussian mixture models (GMM) possess several standard results available in closed form. Some examples utilizing this advantage include the interacting multiple model (IMM) filter and its modifications for tracking maneuvering targets \cite{best1997new}, the multiple hypothesis tracker (MHT) for multiple target tracking \cite{bar2004estimation}, Gaussian sum filter \cite{alspach1972nonlinear} and the probability hypothesis density (PHD) filter \cite{clark2006gm}. These filters are extensively implemented owing to their performance in realistic situations. However, the fusion of such cross-correlated mixture-modelled tracks becomes infeasible due to the unavailability of  closed-form expressions. For instance, decorrelation requires division of prior and posterior densities, for which an exact result is impossible if Gaussian mixtures are involved \cite{ahmed2015s}. Chernoff fusion requires computing non-integer powers of a Gaussian mixture which does not assume a closed-form \cite{gunay2016chernoff}. The arithmetic averaging, though feasible, presents a relatively larger uncertainty due to amalgamation of densities. Therefore, further processing in the form of mode merging and pruning is needed \cite{li2020arithmetic}. Due to such reasons, fusion of Gaussian mixtures is non-trivial in practical scenarios, and we need to search in alternative directions. This article proposes a method which can handle uni-modal and multi-modal Gaussian densities without altering the framework. 
	
	Research on Gaussian mixture fusion is contemporary and various solutions are available in literature. \cite{noack2014nonlinear} addressed the issue of deriving the dependency relation between Gaussian mixture estimates and found that the dependency is itself a Gaussian mixture. No remarks on how to ease the fusion, however, were made. A tracklet \cite{drummond2002track} based fusion strategy for multiple model tracker is proposed in \cite{acar2020decorrelation} where the information decorrelation approach was extended to Gaussian mixture densities and then naive fusion was applied on the decorrelated mixture. This work required division by a Gaussian mixture wherein, a crude approximation of component-wise division was applied. In \cite{visina2019demand}, the dynamic state of an IMM running fusion center was derived when the local tracks are in turn IMMs using the ``inside information". But, this approach was based on the assumption that individual models in the IMM are entirely the same except for the process noise. Thus, the approach is not useful in scenarios where, for e.g. nearly-constant velocity (NCV) and nearly-constant acceleration (NCA) models are employed in parallel.  
	
	Recently, conservative fusion methodology has taken a different direction where the use of Fr\'etchet-means \cite{nielsen2013matrix} are explored. These works are based on the fact that averaging rules are immune to double counting, thus imparting robustness with respect to correlations involved without requiring any extra information from the local trackers. In \cite{ajgl2014conservativeness}, a formal definition of conservativeness was presented and proved for arithmetic and geometric averaging. The definition was based on the amount of uncertainty in the distribution, thus using measures such as the entropy and the Kullback-Leibler divergence. In \cite{li2020arithmetic} the arithmetic averaging was applied on a multitarget tracking scenario using the multi-Bernoulli process in which each component was represented by a Gaussian mixture. In \cite{bailey2012conservative}, the notion conservativeness was examined using the notion of entropy. It was proved that geometric mean fusion increases the entropy for positive values of the exponent, indirectly proving conserativeness.
	
	The generalized framework for Chernoff fusion was proposed in \cite{mahler2000optimal} by Mahler as the generalized Uhlmann-Julier covariance intersection. Also, in \cite{hurley2002information}, Hurley succinctly expressed the relation between Chernoff information and covariance intersection. In \cite{battistelli2014kullback}, the Chernoff fusion was studied for non-Gaussian distribution and was termed the geometric mean density (GMD). Since the covariance intersection (CI) assumes a closed form for Gaussian densities, component-wise CI can be used while fusing Gaussian mixture using the Chernoff information. This was the approach used in \cite{upcroft2005rich}, where the GMM were fused for a bearing-only tracking scenario. A better result was discussed in \cite{julier2006empirical}, where a first order approximation of the power of a Gaussian mixture was employed. This was the first paper that dealt with approximating the power of a Gaussian mixture model with another Gaussian mixture. The result was succeeded by Gunay et al., which employed sigma-points for generalizing the power of a Gaussian mixture \cite{gunay2016chernoff} but it was computationally demanding. Moreover, such a fusion strategy is affected adversely when the individual modes of a mixture are very close which is usually the case with an IMM tracker.  A review of various methods for approximating the non-integer power of a Gaussian mixture was presented in \cite{ajgl2015approximation} along with a comparison.   
	
	This paper presents the development of a new mean based fusion strategy that can deal with Gaussian mixtures with a minimal approximation. For Gaussian mixture densities, the common information takes the form of a mixture density based on the fusion weights ($\omega$). We approximate the common information as Gaussian and plug it into the harmonic mean formulation. The resulting density is available in closed-form. In terms of computation, the approach performs similar to the Chernoff fusion of Gaussian mixtures (which is also the fastest method) and can be shown to be more accurate in terms of root-mean-square error (RMSE).
	 
	Other contributions of the paper are enumerated as follows:
	\begin{enumerate}[1.]
		\item Detailed investigation of existing average based fusion algorithms, specifically the arithmetic mean density (AMD) and the geometric mean density (GMD). Their issues in implementation while addressing T2TF with Gaussian mixture densities are addressed.
		\item The mathematical development of the proposed harmonic mean density based fusion is elucidated along with its various properties in line with the other average based consensus methods.
		\item A comparison among fusion strategies based on two real-life target tracking scenario has been made.
	\end{enumerate}
	
	The rest of the paper is organized as follows; in Section \ref{probForm}, a brief summary of the issue of cross-correlation in a distributed sensor fusion scenario is presented. In Section \ref{review}, a review of  the arithmetic and geometric mean fusion along with their implementation for Gaussian mixtures is presented. These include the covariance intersection, pseudo-Chernoff fusion (PCF) for the geometric averaging, and arithmetic averaging. The proposed harmonic mean density fusion is discussed in Section \ref{proposed}. 
	Section \ref{implementation} presents the approximate implementation strategies for the harmonic mean fusion as it requires division by a Gaussian mixture. The algorithm is simulated on 2D and 3D scenarios, and the on comparative performance of different averaging strategies are presented in Section \ref{simulation}. Finally, the article is concluded in Section \ref{conclusion}.

	\section{Problem Formulation} \label{probForm}
	Considering a scenario with two sensor platforms -- $\mathcal{S}^1$ and $\mathcal{S}^2$ generating processed track density $ p(\mathbf{x}_k|\mathbf{z}_k^i)$ conditioned on current measurement $\mathbf{z}^i_k$, where $k$ is the time step and $i$ $\in$ $\{1,2\}$. 
	Assume that both tracks are conditioned on a common prior at time $k$ which is quantified as $p_c(\mathbf{x}_k) = p(\mathbf{x}_k|\mathbf{z}_{1:k-1}^i) $. Then, 
	\begin{align}
		p(\mathbf{x}_k|\mathbf{z}_{1:k}^i) \propto p(\mathbf{z}^i_k|\mathbf{x}_k) p_c(\mathbf{x}_k).
	\end{align}
	Using naive fusion with such local density results in equation \eqref{naive}
	\begin{align}
		p_N(\mathbf{x}_k|\mathbf{z}_{1:k}^1 \cup \mathbf{z}_{1:k}^2) \propto p(\mathbf{z}^1_k|\mathbf{x}_k)p_c(\mathbf{x}_k)p(\mathbf{z}^2_k|\mathbf{x}_k)p_c(\mathbf{x}_k),\label{naive}
	\end{align}
	where $p_N(\mathbf{x}_k|z_{1:k}^1 \cup z_{1:k}^2)$ is the naively fused density. Thus, the common information is accounted twice which will lead to an optimistic estimate (for e.g. multiplying two Gaussian densities results in a Gaussian density with a lower or same covariance). A heuristic approach would be to use naive fusion while dividing once with the common information,
	\begin{align}
		p(\mathbf{x}_k|\mathbf{z}_{1:k}^1 \cup \mathbf{z}_{1:k}^2) &\propto \frac{p(\mathbf{z}^1_k|\mathbf{x}_k)p_c(\mathbf{x}_k)p(\mathbf{z}^2_k|\mathbf{x}_k)p_c(\mathbf{x}_k)}{p_c(\mathbf{x}_k)},\notag\\
		& \propto \frac{p(\mathbf{x}_k|\mathbf{z}_{1:k}^1)p(\mathbf{x}_k|\mathbf{z}_{1:k}^2)}{p(\mathbf{x}_k|\mathbf{z}^1_k\cap \mathbf{z}^2_k)}.\label{exactBayesian}
	\end{align}
	
	This is also the exact Bayesian formulation for distributed data fusion \cite{lu2019distributed}, where $p(\mathbf{x}_k|z^1_k\cap z^2_k)$ accounts for the density due to common prior. For Gaussian densities, an exact formulation of common prior was presented in \cite{bar1981track} which will require extra information from local trackers. This paper focuses on a class of solutions where the common information can be accounted for (in a sub-optimal sense) by only using the information available. Thus,
	\begin{align}
		p(\mathbf{x}_k|\mathbf{z}^1_k\cap \mathbf{z}^2_k) \propto \mathcal{F}\left(p(\mathbf{x}_k|\mathbf{z}_k^1), p(\mathbf{x}_k|\mathbf{z}_k^2), \omega\right),
	\end{align}
	where,  $\mathcal{F}()$ is an appropriate functional operating on probability density functions, and $\omega$ $\in$ $\{0,1\}$ is a scalar parameter that can be evaluated such that the resulting density is optimal in some sense. Usually, the optimality of $\omega$ is to minimize the trace/determinant of the covariance of the resulting fused density.
	
	Say the local platform employs an IMM tracker, then the posterior density $p(\mathbf{x}_k|z_{1:k}^i)$ consists of estimates from multiple modes packed in a Gaussian mixture  of the form,
	\begin{align}
		p(\mathbf{x}_k|\mathbf{z}_{1:k}^i) = \sum_{m = 1}^{M} \mu^m \mathcal{N}\left(\mathbf{x}_k; \hat{\mathbf{x}}^m_k, \Gamma^m_k\right),
	\end{align}
	where $M$ is the total number of modes, $\mu^m$ is the mode probability of the $m^{\text{th}}$ Gaussian component. An exact solution to the fusion of such tracks is infeasible due to,
	\begin{enumerate}[1.]
		\item Communication constraint as $M\times \mathcal{S}$ number of filter gains and measurement Jacobian matrices need to be communicated at each fusion instant.
		\item Division by a Gaussian mixture \cite{ahmed2015s}.
	\end{enumerate}	
	The second point leads to sub-optimality even if exact procedures are followed. This necessitates the search for robust and sub-optimal class of strategies.

	\section{Review of Arithmetic and Geometric Mean Density}\label{review}
	
	A brief review of methods involving approximating a conservative density using the abstract means \cite{nielsen2019jensen} or the generalized quasi-arithmetic mean \cite{niculescu2006convex} is discussed in this section. As provided in \cite{nielsen2019jensen}, a mixture density $\mathbf{M}_\omega\{p_1(\mathbf{x}_k),p_2(\mathbf{x}_k)\}$ is an M-mixture if it generates an interpolation between the densities $p_1(\mathbf{x}_k)$ and $p_2(\mathbf{x}_k)$ with respect to an $\omega$-weighted mean $\mathcal{M}$, 
	%
	\begin{align}
		\mathbf{M}_\omega\{p_1(\mathbf{x}_k),p_2(\mathbf{x}_k)\} = \frac{\mathcal{M}_\omega\{p_1(\mathbf{x}_k), p_2(\mathbf{x}_k)\}}{\zeta_{\mathcal{M}}\{p_1(\mathbf{x}_k), p_2(\mathbf{x}_k)\}}, \label{M-mixture}
	\end{align} 
	where the normalization constant $\zeta_{\mathcal{M}}\{p_1(\mathbf{x}_k), p_2(\mathbf{x}_k)\}$ ensures that the mixture $\mathbf{M}_\omega(.)$ is a valid probability density function. The mean $\mathcal{M}_\omega$ can be generalized to $n$ number of densities if $\omega$ belongs to a $n$-dimensional probability simplex. Some implementations already existing in the literature are,
	\begin{align}
		\mathcal{M}^g_\omega\{p_1(\mathbf{x}_k), p_2(\mathbf{x}_k) \} &= \left\{p_1(\mathbf{x}_k)\right\}^\omega \left\{p_2(\mathbf{x}_k)\right\}^{1-\omega},\label{GM}\\
		\mathcal{M}^a_\omega\{p_1(\mathbf{x}_k), p_2(\mathbf{x}_k) \} &= \omega p_1(\mathbf{x}_k) + \left(1-\omega\right)p_2(\mathbf{x}_k). \label{AM}
	\end{align}
	Equation \eqref{GM} is the well known geometric mean \cite{julier2006empirical,nielsen2009statistical}, whereas equation \eqref{AM} assumes the form of the arithmetic mean \cite{li2020arithmetic,mclachlan2019finite}.
	
	\subsection{Geometric Mean Density (GMD) Fusion}
\begin{figure}[h!]
		\centering
		\includegraphics[width = 0.95\columnwidth]{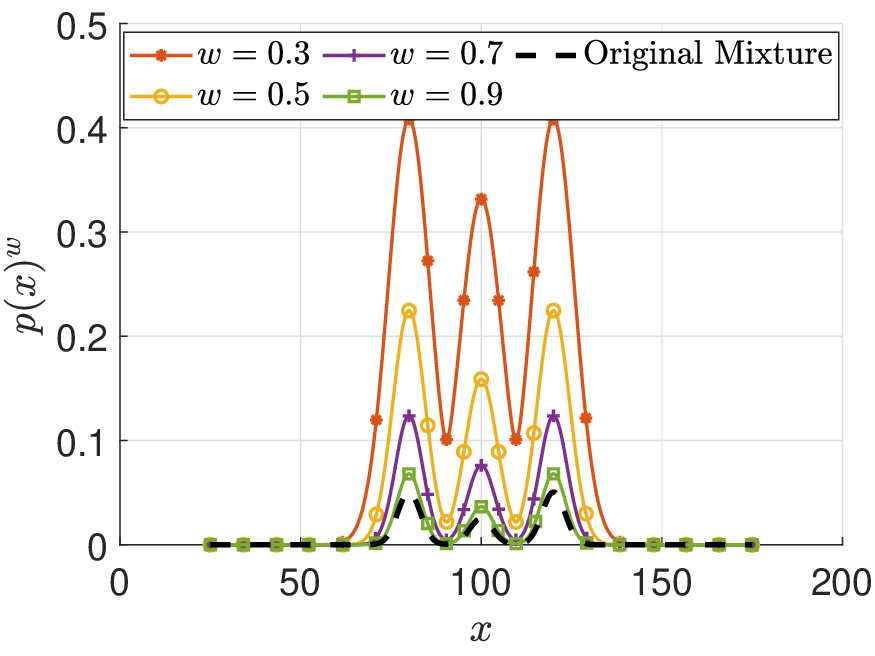}
		\caption{A non-normalized Gaussian mixture raised to a non-integer power $w \in [0,1]$. The original mixture is shown with dashed line, as $\omega$ approaches 0, the density becomes relatively more diffused. } \label{gmmRaised2w}
	\end{figure}

	The geometric mean density (GMD) as proposed in \cite{hurley2002information,mahler2000optimal} is a fusion strategy which assumes the form,
	\begin{align}
		\mathbf{M}^{g}_\omega \{p_1(\mathbf{x}_k),p_2(\mathbf{x}_k)\} = \frac{\left\{ p^{1}(\mathbf{x}_k)\right\}^\omega\left\{p^{2}(\mathbf{x}_k)\right\}^{1-\omega}}{\int \left\{p^{1}(\mathbf{x}_k)\right\}^\omega\left\{p^{2}(\mathbf{x}_k)\right\}^{1-\omega}d\mathbf{x}_k}, \label{gmd}
	\end{align}
	where $\omega\in\left[0,1\right]$ which is selected by solving a suitable optimization problem \cite{uney2011information}. It has been proved in \cite{bailey2012conservative, julier2008fusion} that the fusion strategy in equation \eqref{gmd} is immune to double counting. An appealing property of equation \eqref{gmd} is that it admits a closed form solution for exponential family of distributions. For instance, if the densities $p^{1}(\mathbf{x})$ and $p^{2}(\mathbf{x})$ follow Gaussian distribution,
	\begin{subequations}
		\begin{align}
			p^{1}(\mathbf{x}_k) &\sim \mathcal{N}(\mathbf{x}_k;\hat{\mathbf{x}}_k^1,\Gamma_k^1),\\
			p^{2}(\mathbf{x}_k) &\sim \mathcal{N}(\mathbf{x}_k;\hat{\mathbf{x}}_k^2,\Gamma_k^2),
		\end{align}
	\end{subequations}
	then the geometric mean density fusion reduces to \textit{covariance intersection}, with the fused mean $\hat{\mathbf{x}}^f$ and covariance $\Gamma^f$,
	\begin{subequations} \label{CI}
		\begin{align}
			\Gamma_k^f &= \left( {\omega\Gamma_k^1}^{-1} + {(1 - \omega)\Gamma_k^2}^{-1} \right) ^{-1},\\
			\hat{\mathbf{x}}_k^f &= \Gamma_k^f\left( \omega {\Gamma_k^1}^{-1}\hat{\mathbf{x}}_k^1 + {(1 - \omega)\Gamma_k^2}^{-1}\hat{\mathbf{x}}_k^2 \right).	
		\end{align}
	\end{subequations}
	Though the closed form solutions for exponential families are readily available. The complexity of using GMD for Gaussian mixtures is a harder problem since the non-integer power of a Gaussian mixtures is not always a valid distribution, let alone a closed form availability (also, see Fig. \ref{gmmRaised2w}). The authors in \cite{ahmed2015s} focused solely on methods involving {\color{black}non-integer power of Gaussian mixtures}. But, a simple solution termed as the pseudo-Chernoff fusion uses the following first order approximation for the power of a Gaussian mixture,
	\begin{align}
		\left[p(\mathbf{x}_k)\right]^\omega &= \left[	\sum_{m = 1}^{M} \mu^m \mathcal{N}\left(\mathbf{x}_k; \hat{\mathbf{x}}^m_k, \Gamma^m_k\right)\right]^\omega, \notag \\
		&\approx \sum_{m = 1}^{M} (\mu^m)^\omega \mathcal{N}\left(\mathbf{x}_k; \hat{\mathbf{x}}^m_k, \Gamma^m_k\right)^\omega \label{pcf},
	\end{align} 
	and the power of a Gaussian distribution is a scaled Gaussian distribution with the suitable normalization constant, i.e.
	\begin{align}
		\mathcal{N}\left(\mathbf{x}_k; \hat{\mathbf{x}}^m_k, \Gamma^m_k\right)^\omega \triangleq \alpha_\omega \mathcal{N}\left(\mathbf{x}_k; \hat{\mathbf{x}}^m_k, \frac{ \Gamma^m_k}{\omega}\right),
	\end{align}
	where the normalization constant, $\alpha_\omega$ is,
	\begin{align}
		\alpha_\omega = \sqrt{\frac{\lvert 2\pi\frac{\Gamma^m_k}{\omega}\rvert} {\lvert2\pi\Gamma^m_k  \rvert^\omega} }.
	\end{align}
	Note that this approximation for power of a Gaussian mixture is good only when the components are sufficiently separated and the quality of the approximation deteriorate as the components of the mixture become close to each other \cite{gunay2016chernoff}. Due to this reason, the pseudo-Chernoff fusion is not a decent candidate for IMM trackers especially if modes are based on similar state dynamics. 
	
	Another approximation for the power of a Gaussian mixture known as the sigma-point Chernoff fusion (SPCF) was presented in \cite{gunay2016chernoff} which was based on approximating the powered Gaussian as follows
	\begin{subequations}
		\begin{align}
			q(\mathbf{x}_k) &= \left[	\sum_{m = 1}^{M} \mu^m \mathcal{N}\left(\mathbf{x}_k; \hat{\mathbf{x}}^m_k, \Gamma^m_k\right)\right]^\omega, \\
			&\approx  \sum_{m = 1}^{M}\beta_k^m \mathcal{N}\left(\mathbf{x}_k; \hat{\mathbf{x}}^m_k, \frac{\Gamma^m_k}{\omega}\right).
		\end{align}
	\end{subequations}
	The approximation is appropriate as shown in Fig. \ref{gmmRaised2w}. The modes of the resulting un-normalized density are unchanged, but the variance of each component is inflated which is assumed proportional to $\omega$ in the approximation. The next step is to solve the following optimization problem using deterministic sampling to find the component weights $\beta_k = \begin{bmatrix}\beta_k^1 & \beta_k^2 &\cdots& \beta_k^m  \end{bmatrix}^T$,
	\begin{subequations}\label{optimSP}
		\begin{align}
			\beta_k^* = &\argmin_{\beta_k}\int \left( q(\mathbf{x}_k) - (p(\mathbf{x}_k))^\omega  \right)^2p(\mathbf{x}_k)d\mathbf{x}_k \\
			& \text{s.t. } \beta_k^m \geq 0, \qquad m = 1,\dots, M.	
		\end{align}
	\end{subequations}
	Thus, prior to the fusion of track densities, it is vital to evaluate sigma-points and solve equation \eqref{optimSP} for each density which increases the computation of the fusion process drastically. For the example in \cite{gunay2016chernoff}, the sigma-point based approximation was six times more computationally costly than the approximation in equation \eqref{pcf}. Once the power of the track densities are calculated, one can simply use naive fusion since the independence has been accounted for using $\omega$.
	
	\subsection{Arithmetic Mean Density (AMD) Fusion}
	The arithmetic mean density is the result of applying linear fusion approach to density functions. The AMD fusion formula is given by
	\begin{align} \label{AMD}
		\mathbf{M}^{a}_\omega\left\{p_1(\mathbf{x}_k),p_2(\mathbf{x}_k),\dots, p_n(\mathbf{x}_k)\right\} = \sum_i^N\omega_ip_i(\mathbf{x}_k),
	\end{align}
	which always results in a valid probability density function as long as $\sum\limits_i\omega_i = 1$. Different probability densities can be conveniently fused using the arithmetic average owing to its attractive property of resulting in a closed form mixture as demonstrated in \cite{da2019kullback}. It was also pointed out that AMD also serves as the ``minimizing density" for the average K-L divergence $\mathbb{D}_{KL}$,
	\begin{align}
		\mathbf{M}^{a}_\omega(.) = \argmin_{g(\mathbf{x}_k)} \sum_i\omega_i \mathbb{D}_{KL}\left(p_i(\mathbf{x}_k):g(\mathbf{x}_k)\right),
	\end{align}
	where $g(\mathbf{x}_k)$ is the argument of minimization. The above property does not emphasize on the quality of fusion strategy but can be exploited as an optimization problem that results in $\omega_i$. 
	{\color{black}The minimization merely states that the output (fused) density after AMD fusion is at minimum distance, in the sense of K-L divergence, with the participating local densities. In other words, while each local posterior is replaced by a conservative version of itself, and then fused, the change it undergoes is minimum. This minimization does not provides any motivation for optimal distributed fusion.}
	
	An important property of arithmetic average is that the fusion approach is consistent even if one of the densities are inconsistent \cite{bailey2012conservative}. The GMD in equation \eqref{gmd} however requires that both probability densities are consistent to output a consistent fused result. The conservativeness of AMD has also been proved in \cite{ajgl2014conservativeness}. 
	An implementation of AMD in multisensor-multitarget tracking scenario was performed in \cite{li2020arithmetic} where the linear fusion approach was applied as a target-wise fusion rule to a multi-Bernoulli process. 
	
	A major problem with AMD fusion is its \textit{over-conservativeness}. It can be seen that since AMD fusion is an aggregating process, it results in a much larger variance than either components. This \textit{increased bandwidth} can be advantageous in situations with biased estimators which has an offset with estimator mean. The immunity 
	of AMD to handle offsets and biases can also be extended to cases with misdetection and false alarms thereby increasing its accuracy in such cases. Another advantage of AMD is its computational efficiency for exact formulation of Gaussian mixtures/particles enabling real-time processing. Also note, that the calculation of 
	the normalization constant is not required in this case unlike GMD.
	
	However, as we will see in Section \ref{simulation}, the AMD performs poorly in cases involving unobservable local nodes where feedback is required. The large variance of the AMD leads to a more uncertain fused estimate and usually takes a longer time to settle down. The requirement of clustering/pruning/merging as a post processing step in AMD is also a fundamental disadvantage \cite{li2020arithmetic}.
	
	\section{Proposed Harmonic Mean Density}\label{proposed}
	The generalization of the harmonic mean and their ubiquity is given in \cite{niculescu2006convex, kolmogorov1991selected,  de2016mean}. The weighted harmonic mean of two scalar estimates $\hat{x}^1$ and $\hat{x}^2$ is given by the expression,
	\begin{align}
		H_\omega(\hat{x}_k^1, \hat{x}_k^2) = \frac{\hat{x}_k^1\hat{x}_k^2}{(1-\omega)\hat{x}_k^1 + \omega\hat{x}_k^2}.
	\end{align}
	The harmonic mean can also be obtained from quasi-arithmetic mean using the monotone function $f(x) = \frac{1}{x} $ \cite{bibby1974axiomatisations}. Based on the statistical M-mixture definition given in equation \eqref{M-mixture}, the harmonic mean interpolated between $p_1(\mathbf{x}_k)$ and $p_2(\mathbf{x}_k)$ is,
	\begin{align}
		\mathcal{M}^h_\omega\{p_1(\mathbf{x}_k), p_2(\mathbf{x}_k) \} = \frac{p_1(\mathbf{x}_k)p_2(\mathbf{x}_k)}{(1-\omega)p_1(\mathbf{x}_k) + \omega p_2(\mathbf{x}_k)}, \label{harmonic}
	\end{align}
	and the corresponding density,
	\begin{align}
		\mathbf{M}^h_\omega\{p_1(\mathbf{x}_k), p_2(\mathbf{x}_k) \} = \frac{1}{\zeta^h_{\mathcal{M}}}\frac{p_1(\mathbf{x}_k)p_2(\mathbf{x}_k)}{(1-\omega)p_1(\mathbf{x}_k) + \omega p_2(\mathbf{x}_k)},
	\end{align}
	where,
	\begin{align}
		\zeta^h_{\mathcal{M}} = \int_{\mathbb{R}_n}\frac{p_1(\mathbf{x}_k)p_2(\mathbf{x}_k)}{(1-\omega)p_1(\mathbf{x}_k) + \omega p_2(\mathbf{x}_k)}d\mathbf{x}_k. \label{harmonicNormConst}
	\end{align}
	Comparing equation \eqref{harmonicNormConst} with the exact Bayesian density in equation \eqref{exactBayesian}, the only difference is that the harmonic averaging accounts for the common information $p_c(x)$ as a convex combination of the individual densities (precisely, a weighted arithmetic mean),
	\begin{align}
		p(\mathbf{x}_k|\mathbf{z}^1_k\cap \mathbf{z}^2_k) \propto (1 -\omega)p_1(\mathbf{x}_k|\mathbf{z}^1_k) + \omega p_2(\mathbf{x}_k|\mathbf{z}^2_k).
	\end{align}
	This is also analogous to the geometric mean density $\mathcal{M}^g_\omega$ as,
	\begin{align}
		\mathcal{M}^g_\omega &\propto p_1(\mathbf{x}_k)^\omega p_2(\mathbf{x}_k)^{(1-\omega)}\\
		&= \frac{p_1(\mathbf{x}_k)  p_2(\mathbf{x}_k)}{p_1(\mathbf{x}_k)^{(1-\omega)}p_2(\mathbf{x}_k)^\omega}. 
	\end{align}
	In this case, the common information is approximated by the geometric mean of individual densities. Interestingly, the similarity is shown by the arithmetic average as well, where,
	\begin{align}
		p(\mathbf{x}_k|\mathbf{z}^1_k\cup \mathbf{z}^2_k) &= \omega p_1(\mathbf{x}) + (1-\omega)p_2(\mathbf{x})\\
		&= 	\frac{p_1(\mathbf{x}_k)\times  p_2(\mathbf{x}_k)}{\frac{p_1(\mathbf{x}_k) p_2(\mathbf{x}_k)}{\omega p_1(\mathbf{x}_k) + (1 - \omega) p_2(\mathbf{x}_k)}}. \label{geometricCommonInfo}
	\end{align}
	Hence, in this case,
	\begin{align}
		p(\mathbf{x}_k|\mathbf{z}^1_k\cap \mathbf{z}^2_k) \propto \frac{p_1(\mathbf{x}_k) p_2(\mathbf{x}_k)}{\omega p_1(\mathbf{x}_k) + (1 - \omega) p_2(\mathbf{x}_k)}.
	\end{align}
	Thus, to summarize;
	\begin{enumerate}[1.]
		\item The AMD fusion approximates the common information as the harmonic mean of individual posterior densities.
		\item The GMD uses the geometric mean itself as the common information density.
		\item {\color{black}The HMD approximates the common information as an arithmetic average of the posterior track densities.}
	\end{enumerate}	
	
	\subsection{Properties of Harmonic Average Density}

	\vspace{1ex}
	\begin{theorem}
		The harmonic fusion avoids double counting of information.
	\end{theorem}	
	\begin{IEEEproof}
		The proof is similar to that for the other two mean densities \cite{bailey2012conservative} where conditional dependence is assumed. Hence (omitting $k$),
		\begin{align}
			p(\mathbf{z}^i|\mathbf{x}) = p(\mathbf{z}^{i/j}|\mathbf{x})p(\mathbf{z}^i\cap \mathbf{z}^j|\mathbf{x}),  \label{condInd}
		\end{align}
		where, $p(\mathbf{z}^{i/j}|\mathbf{x})$ denotes the exclusive information present with sensor track $i$ relative to $j$ such that $p(\mathbf{z}^{i/j}|\mathbf{x})\cap{p(\mathbf{z}^{j/i}|\mathbf{x})} = \Phi$. Using Bayes' theorem and proceeding with the assumption above,
		\begin{align}
			&p(\mathbf{x}|\mathbf{z}^1 \cup \mathbf{z}^2)\notag\\ &\propto \frac{p(\mathbf{x}|\mathbf{z}^1)p(\mathbf{x}|\mathbf{z}^2)}{(1-\omega)p(\mathbf{x}|\mathbf{z}^1) + \omega p(\mathbf{x}|\mathbf{z}^2)} \notag\\
			&\propto \frac{p(\mathbf{z}^1|\mathbf{x})p(\mathbf{z}^2|\mathbf{x})p(\mathbf{x})}{(1-\omega)p(\mathbf{z}^1|\mathbf{x}) + \omega p(\mathbf{z}^2|\mathbf{x})}\notag\\
			&={\frac{p(\mathbf{z}^{1/2}|\mathbf{x})p(\mathbf{z}^1\cap \mathbf{z}^2|\mathbf{x}) p(\mathbf{z}^{2/1}|\mathbf{x})p(\mathbf{z}^1\cap \mathbf{z}^2|x)p(\mathbf{x})}{(1-\omega)p(\mathbf{z}^{1/2}|\mathbf{x})p(\mathbf{z}^1\cap \mathbf{z}^2|\mathbf{x}) + \omega p(\mathbf{z}^{2/1})p(\mathbf{z}^1\cap \mathbf{z}^2|\mathbf{x})}}\notag\\
			&= \frac{p(\mathbf{z}^{1/2}|\mathbf{x})p(\mathbf{z}^{2/1}|\mathbf{x})}{(1-\omega)p(\mathbf{z}^{1/2}|\mathbf{x}) + \omega p(\mathbf{z}^{2/1}|\mathbf{x})} p(\mathbf{z}^1\cap \mathbf{z}^2|\mathbf{x})p(\mathbf{x}). \label{doubleCount}
		\end{align}
	\end{IEEEproof}	
	Thus, the term $p(\mathbf{z}^1\cap \mathbf{z}^2|\mathbf{x}) $ which accounts for the `rumor propagation' is only accounted once unlike in equation \eqref{naive}. 	
	\vspace{1ex}

	The proof that the normalization constant $\zeta^h_{\mathcal{M}}$ is convex with respect to $\omega$, similar to that in GMD \cite{bailey2012conservative} is presented below. 	
	\vspace{1ex}
	\begin{theorem}\label{theorem_normConst}
		The normalization constant $\zeta^h_{\mathcal{M}}\{p_1(\mathbf{x}_k), p_2(\mathbf{x}_k)\}$ in equation \eqref{harmonicNormConst} is a convex function of $\omega$ and is less than $1$ for $0\leq\omega\leq1$.
	\end{theorem}	
	\begin{IEEEproof}
		Differentiating equation \eqref{harmonicNormConst} with respect to $\omega$ (omitting $k$),
		\begin{align}
			\frac{\partial \zeta^h_{\mathcal{M}}\{p_1(\mathbf{x}), p_2(\mathbf{x})\}}{\partial\omega} &= \frac{\partial}{\partial\omega}\int_{\mathbb{R}_n}  \frac{p_1(\mathbf{x})p_2(\mathbf{x})}{(1 - \omega)p_1(\mathbf{x}) + \omega p_2(\mathbf{x})}d\mathbf{x} \notag\\
			&= \int_{\mathbb{R}_n} -\frac{p_1(\mathbf{x})p_2(\mathbf{x})\left[p_2(\mathbf{x})-p_1(\mathbf{x})\right]}{\left[(1 - \omega)p_1(\mathbf{x}) + \omega p_2(\mathbf{x}) \right]^2}d\mathbf{x}.\notag
		\end{align}s
		The second derivative is,
		\begin{align}
			\frac{\partial^2 \zeta^h_{\mathcal{M}}\{p_1(\mathbf{x}), p_2(\mathbf{x})\}}{\partial\omega^2} &= \int_{\mathbb{R}_n} 2\frac{p_1(\mathbf{x})p_2(\mathbf{x})\left[p_2(\mathbf{x})-p_1(\mathbf{x})\right]^2}{\left[(1 - \omega)p_1(\mathbf{x}) + \omega p_2(\mathbf{x}) \right]^3}d\mathbf{x}, \notag
		\end{align}
		which is clearly positive for $0\leq\omega\leq1$ since $p_1(\mathbf{x})$ and $p_2(\mathbf{x})$ are valid probability densities. Fig. \ref{normConstants_fig} validates Theorem \ref{theorem_normConst} by showing the comparison between $\zeta^h_{\mathcal{M}}$ and $\zeta^g_{\mathcal{M}}$ v/s $\omega$ for two arbitrarily chosen densities.
		
		\begin{figure}
			\centering
			\includegraphics[width = 0.95\columnwidth]{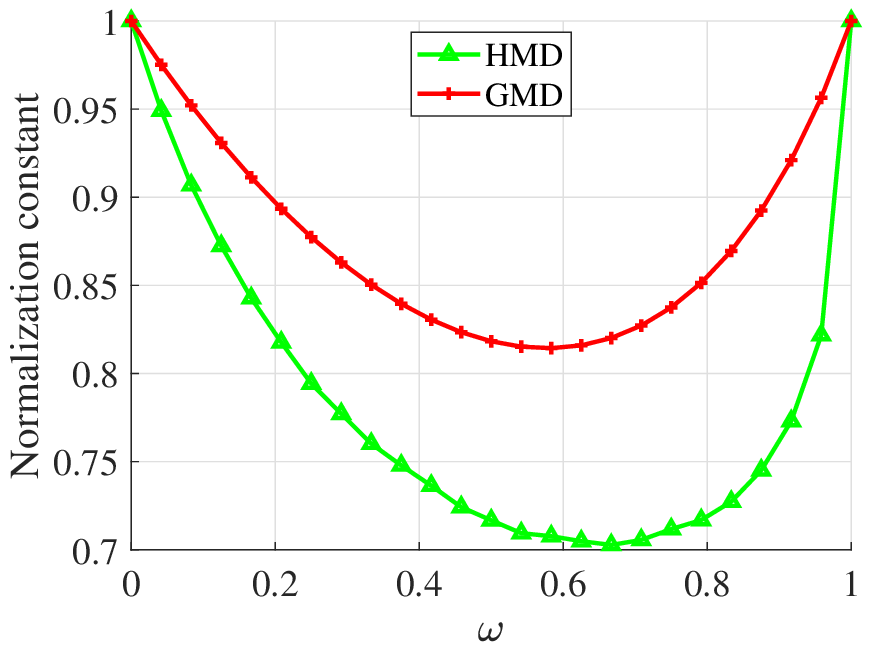}
			\caption{Normalization constant for various values of $\omega$ in case of GMD (red) and HMD (green).}
			\label{normConstants_fig}
		\end{figure}
		It can also be noted that the bounds on $\zeta^h_{\mathcal{M}}$ are on the limiting values of $\omega$,
		\begin{align}
			\zeta^h_{\mathcal{M}} &\leq \left.\zeta^h_{\mathcal{M}} \right|_{\omega \in \{0,1\}} = 1, \label{normConstantHarmonic}
		\end{align}
		since for the case $\omega \in \{0,1\}$, the resulting density reduces to one of the component densities. Consequently, the fused density will always be less diffused, or in the words of Mahler \cite{julier2006empirical}, $\omega$ tends to increase the peakiness of the fused density.
		
	\end{IEEEproof}
	
	\subsection{Harmonic Averaging as a Recursive Fusion Strategy}
	For N number of sensors, the harmonic average assumes the following expression (omitting the argument in $p_i(\mathbf{x}_k)$ for brevity),
	\begin{align}
		\mathcal{M}^h_\omega(p_1,p_2,\dots,p_N) = \frac{\prod\limits_i^N p_i}{\sum\limits_{i}^{N}\omega_i\prod\limits_{\substack{j\\j\neq i}}^N p_j}, \quad \sum\omega_i = 1.
	\end{align}
	However, instead of using the above formulation, a recursive approach to keep the complexity at the minimum can be used. Considering an example with three sensors, the harmonic average is
	\begin{align}
		\mathcal{M}^h_\omega(p_1,p_2,p_3) &= \frac{p_1p_2p_3}{\omega_1p_2p_3 + \omega_2p_1p_3 + \omega_3p_1p_2}, \\
		\frac{1}{\mathcal{M}^h_\omega(p_1,p_2,p_3)}	&= {\frac{\omega_1}{p_1} +\frac{\omega_2}{p_2}+\frac{\omega_3}{p_3}}\notag\\
		&= \frac{1}{\alpha}\left(\frac{\alpha\omega_1}{p_1} +\frac{\alpha\omega_2}{p_2}\right)+\frac{\omega_3}{p_3}\notag\\
		&= \frac{1/\alpha}{\mathcal{M}^h_\omega(p_1,p_2)} + \frac{\omega_3}{p_3}.
	\end{align}
	Here $\sum\limits_i^3\omega_i = 1$ and $\alpha$ is chosen such that $\alpha(\omega_1 + \omega_2) = 1$. Thus it can be shown that,
	\begin{align}
		\mathcal{M}^h_\omega(p_1,p_2,p_3) = \mathcal{M}^h_{\omega,\alpha}\left[\mathcal{M}^h_{\omega,\alpha}(p_1,p_2),p_3\right], \label{recursiveProp}
	\end{align}
	which can be extended to any number of sensors. Therefore, the harmonic averaging can be performed sequentially as long as the resulting density does not diminish, ensuring the amount of certainty in the fusion process. This can also be proved as a property of the harmonic mean density as follows. 
	
	\begin{theorem}
		The harmonic average density is bounded from below by one of the component densities.
	\end{theorem}	
	\begin{IEEEproof}
		The proof is a straightforward property of the harmonic mean with respect to the components. Any abstract mean follows the property \cite{nielsen2019jensen} (omitting $k$),
		\begin{align}
			\inf\{p_1(\mathbf{x}),p_2(\mathbf{x})\}\leq	{\mathcal{M}^h_\omega(p_1(\mathbf{x}),p_2(\mathbf{x}))}\leq \sup\{p_1(\mathbf{x}),p_2(\mathbf{x})\}. \notag 
		\end{align}
		Since it has been proved in equation \eqref{normConstantHarmonic} that the normalization constant is less than 1,
		\begin{align}
			\frac{\mathcal{M}^h_\omega(p_1(\mathbf{x}),p_1(\mathbf{x}))}{\zeta^h_{\mathcal{M}}} \geq {\mathcal{M}^h_\omega(p_1(\mathbf{x}),p_1(\mathbf{x}))} \geq \inf\{p_1(\mathbf{x}),p_2(\mathbf{x})\}. \label{eq_boundHMD}
		\end{align}
		Thus for the normalized pdf, the statement $\inf\{p_1(\mathbf{x}),p_2(\mathbf{x})\}\leq{\mathbf{M}^h_\omega(p_1(\mathbf{x}),p_2(\mathbf{x}))}$ is valid as long as the normalization constant is less than $1$, which is proved in Theorem \ref{theorem_normConst}.
	\end{IEEEproof}

	\subsection{On the K-L Divergence of Harmonic Mean Density}
	\begin{proposition}
		For valid densities $p_1(\mathbf{x})$ and $p_2(\mathbf{x})$, the normalized harmonic density $\mathbf{M}^h_\omega(p_1,p_2)$ should hold,
		\begin{align}
			\mathbb{D}(p_1(\mathbf{x}):\mathbf{M}^h_\omega(p_1,p_2) ) \leq \mathbb{D}(p_1(\mathbf{x}):p_2(\mathbf{x})). 
		\end{align}
		Or conversely,
		\begin{align}
			\mathbb{D}(p_2(\mathbf{x}):\mathbf{M}^h_\omega(p_1,p_2) ) \leq \mathbb{D}(p_2(\mathbf{x}):p_1(\mathbf{x})),
		\end{align}
		where $\mathbb{D}(p:q)$ is the forward Kullback-Leibler divergence between densities $p$ and $q$. Also note that $\mathbb{D}(p:q) \neq \mathbb{D}(q:p)$ in general.
	\end{proposition}
	
	\begin{IEEEproof} 
		 It can be observed that, 
		\begin{align}
			\mathbb{D}(p_1(\mathbf{x}):&\mathbf{M}^h_\omega(p_1,p_2) ) \notag\\&= \notag \int_{\mathbb{R}_n}p_1(\mathbf{x})\ln\frac{\{(1-\omega) p_1(\mathbf{x}) + \omega p_2(\mathbf{x})\}}{p_2(\mathbf{x})} +  \ln(\kappa), \notag\\		
		\end{align}
		where $\kappa$ denotes the normalization constant $\zeta^h_{\mathcal{M}}$. Denoting the cross-entropy between $p_1(\mathbf{x})$ and $p_2(\mathbf{x})$ by $\mathbb{H}_{12}$,
		\begin{align}
			\mathbb{D}(&p_1(\mathbf{x}):\mathbf{M}^h_\omega(p_1,p_2) ) \notag\\	&= \int p_1(\mathbf{x})\ln\{ (1-\omega)p_1(\mathbf{x}) + \omega p_2(\mathbf{x})\} + \mathbb{H}_{12} + \ln(\kappa) \notag\\
			&= 	\mathbb{D}(p_1(\mathbf{x}):p_2(\mathbf{x})) + \int p_1(\mathbf{x})\ln\{(1-\omega) p_1(\mathbf{x}) + \omega p_2(\mathbf{x})\}\notag \\ &\qquad - \int p_1(\mathbf{x})\ln p_1(\mathbf{x}) + \ln(\kappa) \notag \\
			&= \mathbb{D}(p_1(\mathbf{x}):p_2(\mathbf{x})) - \mathbb{D}(p_1(\mathbf{x}):p_{a}(\mathbf{x}))  + \ln(\kappa).
		\end{align}
		Conversely,
		\begin{align}
			\mathbb{D}(&p_2(\mathbf{x}):\mathbf{M}^h_\omega(p_1,p_2) ) \notag\\&= \mathbb{D}(p_2(\mathbf{x}):p_1(\mathbf{x})) - \mathbb{D}(p_2(\mathbf{x}):p_{a}(\mathbf{x})) + \ln(\kappa),
		\end{align}
		where the standard result $ \mathbb{H}_{12} = \mathbb{D}(p_1(\mathbf{x}):p_2(\mathbf{x})) + \mathbb{H}(p_1)$ is used and $p_{a}(\mathbf{x})$ is the arithmetic mean density between $p_1(\mathbf{x})$ and $p_2(\mathbf{x})$.
		
		Note that $\ln(\kappa)$ is less than $0$ owing to equation \eqref{normConstantHarmonic} and $\mathbb{D}(p_2(\mathbf{x}):p_{a}(\mathbf{x}))$ is always positive, thus proving the proposition.
	\end{IEEEproof}
	The statement can be generalized to $n$ number of densities $p_i(\mathbf{x})$ $i = \{1,2, \cdots, n\}  $.
	\begin{align}
		\mathbb{D}(p_i(\mathbf{x}):M^h_\omega(p_1,\dots ,p_n) ) \leq \mathbb{D}(p_i(\mathbf{x}):\max\{p_j(\mathbf{x})\}), 
	\end{align}
	where $j = \{1,2, \cdots, n\}$, $j \neq i$. Note that,
	\begin{align}
		\mathbb{D}(p_i(\mathbf{x}):p_j(\mathbf{x})) \geq \mathbb{D}(p_i(\mathbf{x}):\max\{p_j(\mathbf{x})\}), \text{ for $j\neq i$.}
	\end{align} 
	therefore the harmonic average density is placed in such a way that its Kullback-Leibler divergence with participating densities is less than that among the densities themselves, which is essential for pooling. A major result on divergence minimization of harmonic mean density is also provided in \cite{sharmanhmd}.

	Another interesting property on K-L divergence can be deduced for any arbitrary density $g(\mathbf{x})$ as follows:
	\begin{align}
		\mathbb{D}(g(\mathbf{x}) : \mathbf{M}^h_\omega(p_1,p_2)) &= - \int g(\mathbf{x}) \log\left(\frac{g(\mathbf{x})}{\{(1-\omega) p_1(\mathbf{x}) + \omega p_2(\mathbf{x})\}}\right)d\mathbf{x} + \notag \\ 
				&\int \left[g(\mathbf{x}) \log(g(\mathbf{x})) + g(\mathbf{x})\log\left(\frac{g(\mathbf{x})}{p_1(\mathbf{x})p_2(\mathbf{x})}\right)\right]d\mathbf{x} + \log(\kappa) \notag \\
				&= \sum_{i=1 }^2 \mathbb{D}(g(\mathbf{x}) : p_i(\mathbf{x})) - \mathbb{D}(g(\mathbf{x}) :p_{a}(\mathbf{x}) ) +  \log(\kappa) \notag \\
				&\leq \sum_{i=1 }^2 \mathbb{D}(g(\mathbf{x}) : p_i(\mathbf{x})) - \mathbb{D}(g(\mathbf{x}) :p_{a}(\mathbf{x}) ) \label{eq_dkl_2}
	\end{align}
	Eqn. \eqref{eq_dkl_2} provides an information theoretic viewpoint for the placement of harmonic mean density relative to the true density $g(\mathbf{x})$ and its arithmetic average $p_{a}(\mathbf{x})$.
	
	\subsection{On the Monotonicity of Harmonic Mean Density}
	The shape preserving property of the harmonic average density can be proved using the monotonicity of a function of two variables. A function of two variables $f:(x,y)\in\mathbb{R}^2$ can be termed monotonic if for fixed $(x,y)$ and $(x',y')$
	\begin{align}
		(x\leq x' \text{ and } y\leq y') \implies f(x,y) \leq f(x',y') \label{monotonicProp}
	\end{align}  
	Equation. \eqref{monotonicProp} can be proved for HMD as follows,
	\begin{theorem}
		The harmonic mean density follows the monotonic property in equation \eqref{monotonicProp}.
	\end{theorem}
	\begin{IEEEproof}
		The proof here is shown for two densities $p_1(\mathbf{x})$ and $p_2(\mathbf{x})$ which can be extended to any number of sensors using equation \eqref{recursiveProp}. The gradient of the harmonic average is (omitting the variable $\mathbf{x}$ and $k$ for clarity).
		\begin{align}\label{eq_grad_hmd}
			&\nabla(\mathbf{M}^h_\omega(p_1,p_2)) = \notag\\&\quad\left[\frac{1}{\omega_1p_1^2}\nabla(p_1) + \frac{1}{\omega_2p_2^2}\nabla(p_2) \right](\mathbf{M}^h_\omega(p_1,p_2))^2.
		\end{align} 
		It can be seen that for fixed $p_1(\mathbf{x})$, the harmonic average is monotonic with respect to $p_2(\mathbf{x})$ and monotonic w.r.t. $p_1(\mathbf{x})$ when the other is fixed. Thus from equation \eqref{eq_grad_hmd},
		\begin{subequations} \label{monotonicProof}
			\begin{align}
				\text{if }\tilde{p_1}\leq p_1 &\implies  \mathbf{M}^h_\omega(\tilde{p_1},\gamma) \leq \mathbf{M}^h_\omega(p_1,\gamma),\label{monotonicProof_1}
			\end{align}
			$\text{and,}$
			\begin{align}
				\text{if }\tilde{p_2} \leq p_2 &\implies  \mathbf{M}^h_\omega(\gamma,\tilde{p_2}) \leq \mathbf{M}^h_\omega(\gamma,p_2),\label{monotonicProof_2}
			\end{align}
		\end{subequations}
		where $\gamma$ is an arbitrary constant. The proof of equation \eqref{monotonicProp} for harmonic average is a direct consequence of equations \eqref{monotonicProof_1} and \eqref{monotonicProof_2}.
	\end{IEEEproof}	
	
	\section{Implementation} \label{implementation} 
	The major issue in implementation of HMD is its denominator which takes the form of a valid probability density mixture. Closed form solutions for division by a mixture do not exist \cite{ahmed2015s}, hence approximate solutions are used. 
	
	Considering Gaussian densities, the simplest solution to the division problem is to use a normal approximation for the denominator, 
	\begin{align}
		\omega \mathcal{N}\left(\hat{\mathbf{x}}^1_k, \Gamma^1_k\right)  + (1 - \omega ) \mathcal{N}\left(\hat{\mathbf{x}}^2_k, \Gamma^2_k\right) \approx \mathcal{N}\left(\mathbf{x}_k; \hat{\mathbf{x}}^{eq}_k, \Gamma^{eq}_k\right), \label{eqn_Gauss_approx_den}
	\end{align}
	where $\hat{\mathbf{x}}^{eq}_k$ and $\Gamma^{eq}_k$ are respectively the mean and covariance of the resulting Gaussian approximation of the mixture,
	\begin{align}
		\hat{\mathbf{x}}^{eq}_k &\triangleq \sum_{m = 1}^{M} \mu^m \hat{\mathbf{x}}^m_k \label{GaussApproxMean},\\
		\Gamma^{eq}_k &\triangleq \sum_{m = 1}^{M} \mu^m  \left[\Gamma^m_k + (\hat{\mathbf{x}}^{eq}_k - \hat{\mathbf{x}}^m_k )(\hat{\mathbf{x}}^{eq}_k - \hat{\mathbf{x}}^m_k )^T \right]. \label{GaussApproxCov}
	\end{align}
	Let $p(\mathbf{x}_k|z^i_k) = \mathcal{N}(\mathbf{x}_k; \hat{\mathbf{x}}^{i}_k, \Gamma^{i}_k)$ and $p(\mathbf{x}_k|z^j_k) = \mathcal{N}(\mathbf{x}_k; \hat{\mathbf{x}}^{j}_k, \Gamma^{j}_k)$, then using the Gaussian equivalent for denominator,
	\begin{align}
		p(\mathbf{x}_k|\mathbf{z}^i_k\cup \mathbf{z}^j_k) &= \frac{\mathcal{N}(\mathbf{x}_k; \hat{\mathbf{x}}^{i}_k, \Gamma^{i}_k) \mathcal{N}(\mathbf{x}_k; \hat{\mathbf{x}}^{j}_k, \Gamma^{j}_k) }{\mathcal{N}\left(\mathbf{x}_k; \hat{\mathbf{x}}^{eq}_k, \Gamma^{eq}_k\right)}\\
		&\propto \mathcal{N}\left( \mathbf{x}_k;\hat{\mathbf{x}}^f_k, \Gamma^f_k \right), \label{GaussDivision}
	\end{align}
	where the superscript $f$ stands for \textit{fused}. The fused mean and covariance are then,
	\begin{subequations}\label{HMfused_eq}
		\begin{align} 
			\Gamma^f_k &= \left( \Gamma^{i^{-1}}_k + \Gamma_k^{j^{-1}}  - \Gamma_k^{eq^{-1}}\right)^{-1}\label{fusedMean},\\
			\hat{\mathbf{x}}^f_k &= \Gamma^f_k \left[\left(\Gamma^{i^{-1}}_k\hat{\mathbf{x}}^{i}_k +  \Gamma^{j^{-1}}_k\hat{\mathbf{x}}^{j}_k \right)  -   \Gamma_k^{eq^{-1}}\hat{\mathbf{x}}^{eq}_k\right],\label{fusedCov}
		\end{align}
	\end{subequations}
	where standard results on division and product of Gaussian densities have been used \cite{acar2020decorrelation,bar2004estimation}. $\hat{\mathbf{x}}^{eq}_k$ and $\Gamma^{eq}_k$ are as mentioned in equations \eqref{GaussApproxMean} and \eqref{GaussApproxCov} respectively for weights $\omega$ and $(1 - \omega)$. 
	
	Note that the division of the Gaussian distributions is only valid for certain conditions which is met by HMD and can be proved as follows.
	
	\begin{proposition}
		The division of Gaussian densities as performed in equation \eqref{GaussDivision} is always valid.
	\end{proposition}
	\begin{IEEEproof}
		See \ref{appendix_proof_division}.
	\end{IEEEproof}	
	
	
	{\color{black}
		It is worth noting that the proposed method of implementation of HMD is similar to different extensions of CI with corresponding interpretations for the common information component $(\hat{\mathbf{\gamma}}, \mathbf{\Gamma})$. For instance, the ellipsoidal intersection (EI) provides the fused result at time $k$ as,
		\begin{align}
			\Gamma^f_k &= \left[{\Gamma^i_k}^{-1} + {\Gamma^j_k}^{-1} - {\Gamma^{EI}_k}^{-1}\right]^{-1} \\
			\hat{\mathbf{x}}^f_k &= \Gamma^f_k \left[{\Gamma^i_k}^{-1}\hat{\mathbf{x}}^i_k + {\Gamma^j_k}^{-1}\hat{\mathbf{x}}^j_k - \hat{\mathbf{\gamma}}^{EI}_k{\Gamma^{EI}_k}^{-1} \right]
		\end{align} 
		See \cite{sijs2010state} for the interpretations of $\left(\hat{\mathbf{\gamma}}^{EI}_k, \Gamma^{EI}_k\right)$. Similarly, the inverse covariance intersection (ICI) in \cite{noack2017decentralized} suggests that the common information is a weighted arithmetic average of the components. 
		\begin{align}
			\Gamma^f_k  &= \left[{\Gamma^i_k}^{-1} + {\Gamma^j_k}^{-1} - {\Gamma^{ICI}_k}^{-1}\right]^{-1} \\
			\hat{\mathbf{x}}^f_k &= \Gamma^f_k \bigg[{\Gamma^i_k}^{-1}\hat{\mathbf{x}}^i_k + {\Gamma^j_k}^{-1}\hat{\mathbf{x}}^j_k \notag \\
			&\quad- \left(\omega{\Gamma^{ICI}_k}^{-1}\hat{\mathbf{x}}^i_k + (1-\omega){\Gamma^{ICI}_k}^{-1}\hat{\mathbf{x}}^j_k\right)  \bigg]
		\end{align}
		where, 
		\begin{align}
			\Gamma^{ICI}_k = \omega\Gamma^i_k + (1-\omega)\Gamma^j_k
		\end{align}
		The structure of common information in this case is very similar to the proposed method except the spread of means term in the Gaussian mixture equivalent of Eqn. (57). The difference is due to the fact that HMD uses arithmetic average of local densities rather than that of parameters as in the case of ICI. When fusing estimates with equal means, the proposed method matches exactly with ICI. The interpretation of spread-of-means is discussed elaborately in subsequent section.
		
	}
	The scope of this work is restricted to the use of the Gaussian approximation of the common information, excluding the exact development as a scope for future work.
	{\color{black}
		\paragraph{Implications of the spread of means term in eqn. \eqref{GaussApproxCov}}
		The reader might think that the proposed method of implementation is dubious in the sense that one can increase the spread of means terms by arbitrarily increasing the Euclidean distance between individual means. However, this is not true if the densities are tested for association before fusion. The association test in such case would reject the hypothesis that local densities belong to the same target if normalized distance between the mean is large. For a positive association test, the squared distance between the means has to be less than (scalar case),
		\begin{align}
			(\mu_1 - \mu_2)^2 \leq \gamma_\alpha\times(\sigma_1^2 + \sigma_2^2 - 2\sigma_{12})
		\end{align}
		
		where $\gamma$ is a threshold with $(1-\alpha)$ as the confidence level. $\mu_i$ and $\sigma_i$ are the $i^\text{th}$ mean and standard deviation respectively. $\sigma_{12} = \rho\sqrt{\sigma_1\sigma_2}$ is the correlation coefficient between the two estimates, with $\rho$ as a linear correlation coefficient. Thus, arbitrarily large separation between the individual means is rejected by the test.
		
		Another implication is that if the local densities are assumed consistent, their estimate should correspond to their respective mean square error (MSE). Since both the estimates are representing the same quantity, they should statistically lie within their respective covariance ellipsoids. Thus, local consistency is required for consistency of HMD. 
		
		It should also be noticed that the proposed implementation might result in lower covariance than the local optimum see \cite[see eqn. (8.4.4-5)]{bar1995multitarget}. However, as per the definition in \cite{bailey2012conservative}, a pdf $p(\mathbf{x})$ is inconsistent if at the true value $\mathbf{x}_t$,
		\begin{align}
			p(\mathbf{x}_t) = 0 \label{eq_inconsistent}
		\end{align}
		which would mean that the density is degenerate at the true value. Using naive fusion repetitively leads to a more optimistic covariance, which ultimately leads to the degeneracy depicted in equation \eqref{eq_inconsistent}. Since HMD is a pooled mixture, it enjoys the lower bound in \eqref{eq_boundHMD} and coupled with the fact that it removes double counting of information, it is consistent as long as both the participating densities are consistent. 
	}
	
	It can be observed that using the same formulation allows us to deal with the scenario when track densities are a Gaussian mixture
	\begin{align}
		p(\mathbf{x}_k|\mathbf{z}_k^1) = \sum_i^M \alpha_i\mathcal{N}\left( \mathbf{x}_k;\hat{\mathbf{x}}^{i}_k, \Gamma^{i}_k \right), \notag\\
		p(\mathbf{x}_k|\mathbf{z}_k^2) = \sum_j^N \beta_j\mathcal{N}\left( \mathbf{x}_k;\hat{\mathbf{x}}^{j}_k, \Gamma^{j}_k \right). \notag
	\end{align} 
	Then, the fused density is another Gaussian mixture with the $M\times N$ number of modes and,
	\begin{align}
		p(\mathbf{x}_k|\mathbf{z}^1_k\cup \mathbf{z}^1_k) = \sum_i^M\sum_j^N \alpha_i\beta_j\kappa_{ij}\mathcal{N}\left( \mathbf{x}_k;\hat{\mathbf{x}}^{f}_k, \Gamma^{f}_k \right),
	\end{align}
	where $\hat{\mathbf{x}}^{f}_k$ and $\Gamma^{f}_k$ are exactly the same as in equations \eqref{fusedMean} and \eqref{fusedCov}. {\color{black}The scaling factor $\kappa_{ij}$ is the result of product of Gaussian densities and division by a Gaussian mixture equivalent. It is given by,
		\begin{align}
			\kappa_{ij} = \frac{\mathcal{N}\left(\hat{\mathbf{x}}^j_k; \hat{\mathbf{x}}^i_k, \Gamma^i_k + \Gamma^j_k\right)}{\mathcal{N}\left(\hat{\mathbf{x}}^m_k;\hat{\mathbf{x}}^{\text{naive},ij}_k, \Gamma^m_k-\Gamma_k^{\text{naive},ij} \right) }\times \frac{\abs{\Gamma^m_k-\Gamma_k^{\text{naive},ij}}}{\abs{\Gamma_k^{\text{naive},ij}}}
		\end{align}
		where, $\left(\hat{\mathbf{x}}^{\text{naive},ij}_k,\Gamma_k^{\text{naive},ij}\right)$ is the result of naive fusion between $i^\text{th}$ and $j^\text{th}$ densities. The quantities $\left(\hat{\mathbf{x}}^m_k,\Gamma^m_k\right)$ arise from the Gaussian approximation of the denominator mixture,}
	\begin{align}
		&\omega_1p(\mathbf{x}_k|\mathbf{z}_k^1) + \omega_2p(\mathbf{x}_k|\mathbf{z}_k^2) = \notag\\
		&\omega_1 \sum_i^M \alpha_i\mathcal{N}\left( \mathbf{x}_k;\hat{\mathbf{x}}^{i}_k, \Gamma^{i}_k \right) + \omega_2 \sum_j^N \beta_j\mathcal{N}\left( \mathbf{x}_k;\hat{\mathbf{x}}^{j}_k, \Gamma^{j}_k \right), \notag\\
		& =  \sum_m^{M+N} \mu^m\mathcal{N}\left( \mathbf{x}_k;\hat{\mathbf{x}}^{m}_k, \Gamma^{m}_k \right),\quad\resizebox{0.48\linewidth}{!}{$m = \begin{cases}
				i, \quad 1 \leq m \leq M\\
				j, \quad M+1 \leq m \leq M+N
			\end{cases}$} 
	\end{align}
	where,
	\begin{align*}
		\mu^m = \begin{cases}
			\omega_1\alpha_i,\quad 1 \leq m \leq M\\
			\omega_2\beta_j,\quad  M+1 \leq m \leq M+N.
		\end{cases}
	\end{align*}

		\subsection{An Empirical Study}\label{subsec_empStudy}
\begin{figure}[h!]
			\centering
			\includegraphics[width = \columnwidth]{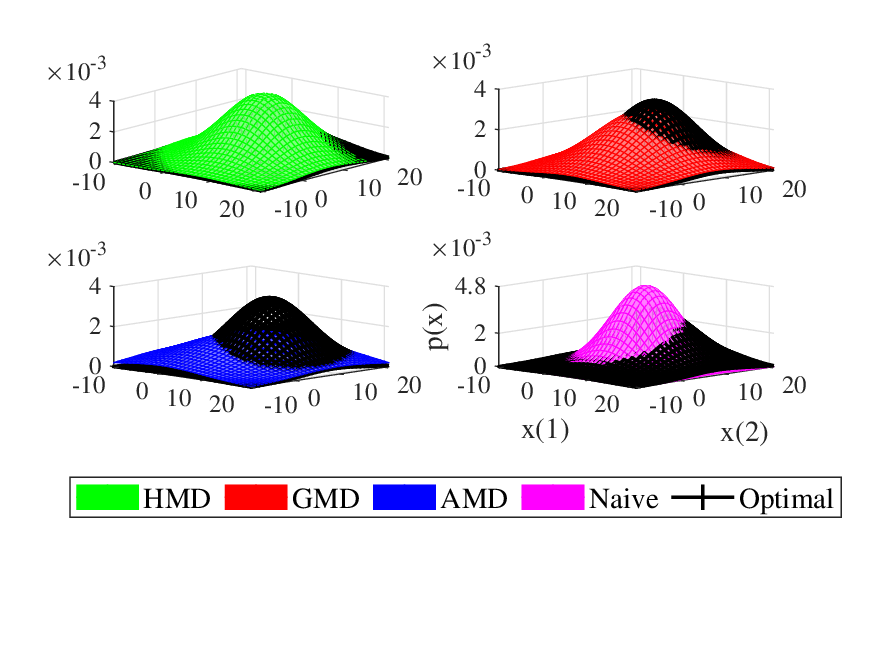}
			\vspace{-1.3cm}
			\caption{Fused density using various methods for a correlation coefficient of $\rho = 0.5$ among local densities.}
			\label{densityFig}
		\end{figure}

		The properties of HMD elucidated in the previous section is observed and compared with the other conservative fusion strategies. Two correlated Gaussian densities $p_1(\mathbf{x})$ and $p_2(\mathbf{x})$ with mean $\mathbf{\hat{x}}_1 = [1,3]^T$ and $\mathbf{\hat{x}}_2 = [7,10]^T$, and covariance $\Gamma_1 = 100\mathbf{I_2}$ and $\Gamma_2 = 50\mathbf{I_2}$ are fused as per the GMD in equation \eqref{CI}, AMD in equation \eqref{AMD} as well as the proposed fusion strategy. The dependence is quantified as,
		\begin{align}
			\Gamma_{12} = \rho\sqrt{\Gamma_1}\sqrt{\Gamma_2},
		\end{align}
		where, $\rho$ is the correlation coefficient taking only positive values between 0 and 1 (tracking applications with homogeneous sensors rarely possess negative correlation for the same target). The optimal density is calculated using the maximum likelihood approach taking into account the correlation matrix $\Gamma_{12}$. 
		The resulting fused density functions are plotted in Fig. \ref{densityFig}. It can be observed that both HMD and GMD are close to the optimal density whereas the arithmetic mean has the largest variance. For different values of $\rho$, the plots of conservative densities will remain the same if $\omega$ is unchanged.
		
\begin{figure}[h!]
			\begin{subfigure}[t]{0.5\columnwidth}
				\centering
				\includegraphics[width = \textwidth]{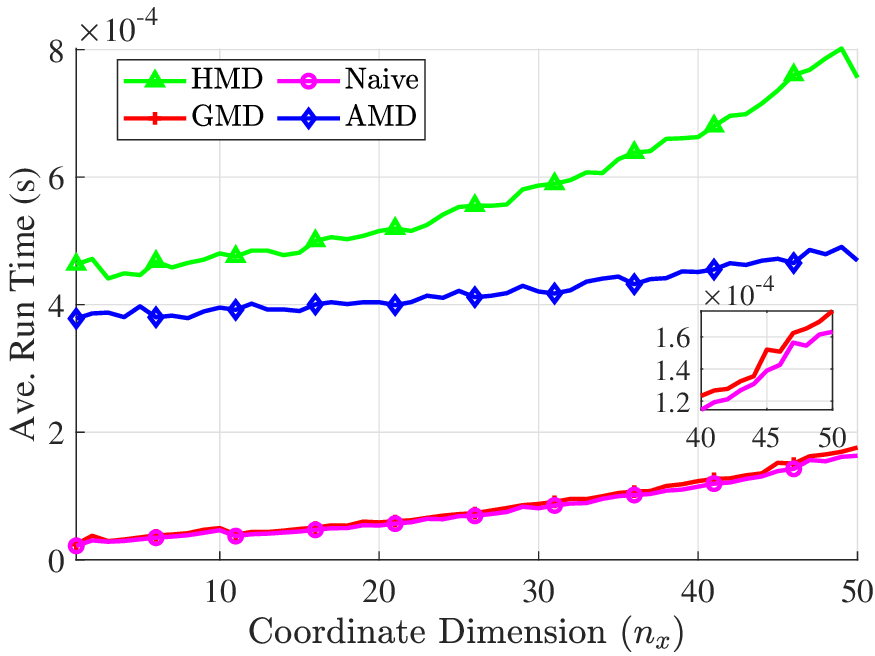}
				\caption{}
				\label{timeGauss_fig}
			\end{subfigure}%
			~
			\begin{subfigure}[t]{0.5\columnwidth}
				\centering
				\includegraphics[width = \textwidth]{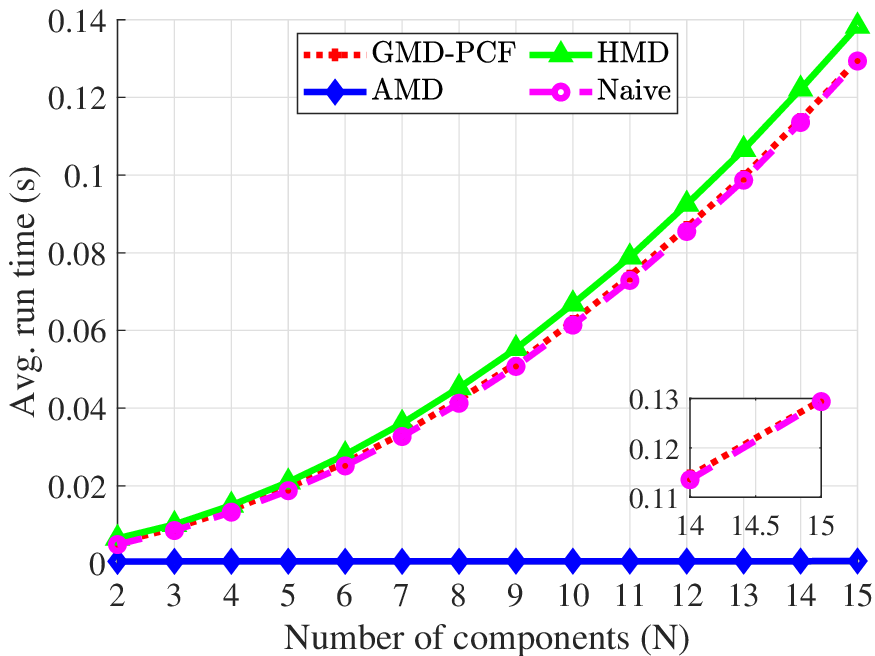}
				\caption{}
				\label{timeGMM_fig}
			\end{subfigure}
			\caption{ \color{black} (a) Average run time v/s dimension of the Gaussian density and, (b) average run time v/s number of components in a  Gaussian mixture.}
			\label{time_fig}
		\end{figure}

		To evaluate the run time of fusion strategies, 
		the dimensions of the Gaussian densities along with the number of components for the Gaussian mixtures are varied. The plots for computation time over 500 Monte-Carlo runs are shown in Fig. \ref{time_fig}. It can be seen in Fig. \ref{timeGauss_fig} 
		that for the Gaussian case, HMD is slightly slower. The GMD and Naive fusion require almost equal computation since both need two inverse operations whereas HMD requires three (for fusing two densities).
		It is also observed that the AMD has a higher average run time in case of Gaussian density, which is due to the fact that overhead of computing Gaussian approximation of the resulting mixture has been added.  Also note that in case of Gaussian mixture in Fig. \ref{timeGMM_fig}, pseudo-Chernoff fusion (PCF) is employed which is the fastest method for fusing Gaussian mixtures using GMD \cite{julier2006empirical} and poses inaccuracies when fusing densities are spatially closer to each other. Also, \cite{gunay2016chernoff} states that the sigma-point approximation based method is six times more computationally costlier than PCF. In either cases, HMD is as computationally efficient as GMD and naive fusion respectively without the requirement of any additional changes to the framework. The results were obtained in MATLAB{\textsuperscript\textregistered}  2020 on a computer with Intel{\textsuperscript\textregistered}  Core{\textsuperscript\texttrademark} i7-9750H CPU @ 2.6 GHz and 16 GB of RAM.
		
		\section{Simulation}\label{simulation} 
		
		To test the performance of the proposed strategy in real-time target tracking environment, two simulation scenarios are presented, one of which is a near constant velocity (NCV) target in 3D, comprising of three sensors and the other is a 2D maneuvering target, tracked using two bearings-only sensors using IMM. Thus, the scenarios cover the fusion of both uni-modal and multi-modal Gaussian densities and are elaborated in different subsections described below. Note that no effort has been made to evaluate the optimal value of the fusion weight $\omega$ and a typical value of $0.5$ is taken. The baseline methods considered for the comparison are :
		\begin{itemize}
			\item Geometric Mean Density (GMD): Covariance intersection formulation for Gaussian densities.
			\item Arithmetic Mean Density (AMD): Weighted amalgamation of local track densities without any pruning/merging.
			\item Centralized EKF: Centralized fusion of measurements using the extended Kalman filter (EKF) used as a lower bound.
		\end{itemize} 
		\subsection{Three Sensor NCV Scenario}
\begin{figure}[h!]
			\centering
			\includegraphics[width = \columnwidth]{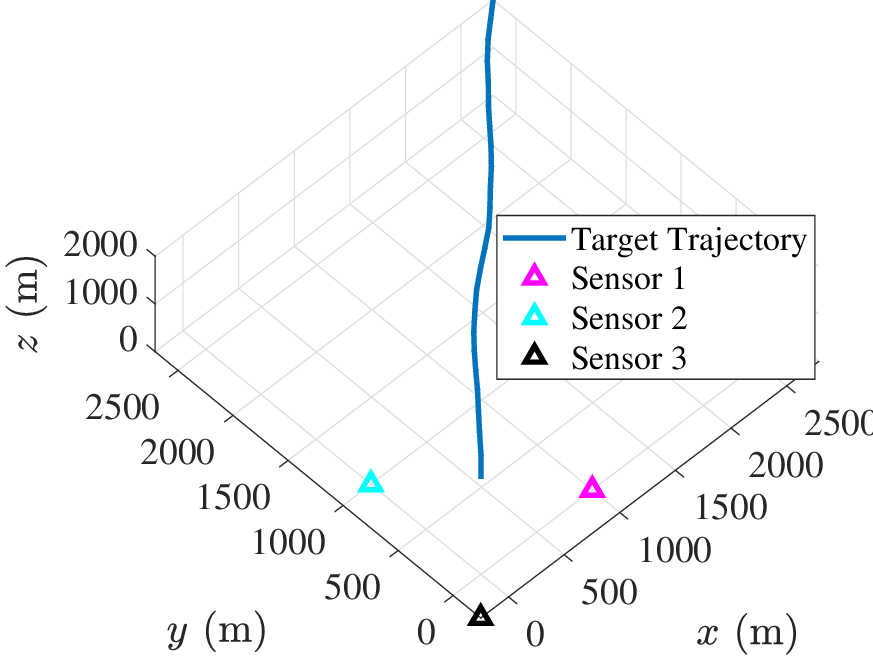}
			\caption{Target and sensor engagements for scenario 1.}
			\label{scenario_1_fig}
		\end{figure}

		In this case, a 3D target trajectory is generated using the near constant velocity motion model with a high process noise intensity, as shown in the Fig. \ref{scenario_1_fig}, where the target starts at the origin at an altitude of $2000$ meters. The state vector $\mathbf{x}_k \in \mathbb{R}^6$ comprises of 3D Cartesian position and velocities, i.e. $\mathbf{x}_k$ = $\begin{bmatrix}x_k&\dot{x}_k&y_k&\dot{y}_k&z_k&\dot{z}_k\end{bmatrix}$. The local nodes employ extended Kalman filter (EKF) with no feedback from the fusion center.
		
		The discrete time target motion and sensors measurements are modeled by the following equations
		\begin{subequations} \label{modelScen_1_eq}
			\begin{align}
				\mathbf{x}_{k+1} &= \mathbf{F}\mathbf{x}_k + \mathbf{w}_k, \\
				\mathbf{z}_k &= \mathbf{h}(x_k,y_k,z_k) + \mathbf{v}_k,
			\end{align}
		\end{subequations}
		where,
		\begin{align*}
			\mathbf{F} &= \begin{bmatrix}
				\mathbf{I}_3 & \Delta T\mathbf{I}_3 \\
				\mathbf{0}_3 & \mathbf{I}_3
			\end{bmatrix} \quad \text{and,} \\
			\mathbf{h}(x_k,y_k,z_k) &= 	\begin{bmatrix} r_k & \theta_k & \phi_k \end{bmatrix}^T,				
		\end{align*}
		where,
		\begin{align*}
			r_k &= \sqrt{x_k^2 + y_k^2 +z_k^2};\quad \theta_k = \tan^{-1}\left(\frac{y_k}{x_k}\right); \\
			\text{and, } \phi_k &= \tan^{-1}\left(\frac{z_k}{\sqrt{x_k^2 + y_k^2}}\right).
		\end{align*}
		$\Delta T$ is the sampling time, and $\mathbf{w}_k$, $\mathbf{v}_k$ are uncorrelated, zero-mean Gaussian distributed process noise vector and measurement noise vector respectively, such that $\forall \{k,j\}$,
		\begin{align*}
			E[\mathbf{w}_k\mathbf{w}_j] = \delta_{kj}\mathbf{Q};\quad E[\mathbf{w}_k\mathbf{v}_j] &= \mathbf{0};\quad E[\mathbf{v}_k\mathbf{v}_j] = \delta_{kj}\mathbf{R}, 
		\end{align*}
		where $\mathbf{Q}$ and $\mathbf{R}$ are the process noise and measurement noise covariance matrix respectively, 
		\begin{align*}
			\mathbf{Q} &= \tilde{q}\begin{bmatrix}
				\frac{\Delta T^3}{3} \mathbf{I}_3 & \frac{\Delta T^2}{2} \mathbf{I}_3 \\
				\frac{\Delta T^2}{2} \mathbf{I}_3 & \Delta T \mathbf{I}_3
			\end{bmatrix}, \\
			\mathbf{R}	&= \text{diag}(\sigma_r^2, \sigma_\theta^2, \sigma_\phi^2),			
		\end{align*}
		and $\tilde{q}$ is the process noise intensity in $\text{meter}^2/\text{second}^3$. The parameters of the scenario are stated in Table \ref{scenario_1_tab}. Note that separate values for process noise intensity in each coordinate have been included. 
		\begin{figure}[h!]
			\centering
			\includegraphics[width = \columnwidth]{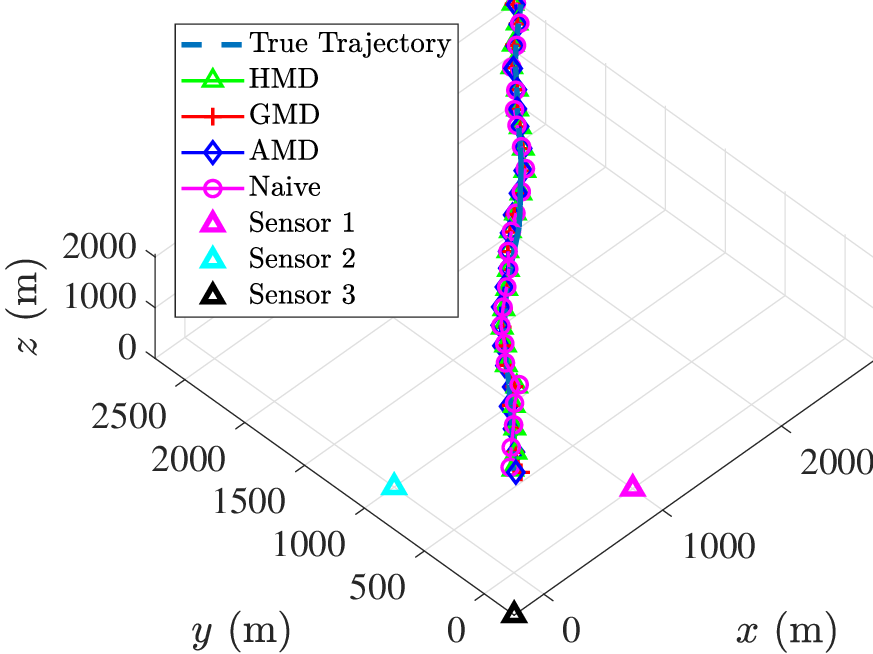}
			\caption{Tracks estimates from various fusion strategies}
			\label{fusedScen_1_fig}
		\end{figure}
		
		\begin{table}[h!]
			\centering
			\caption{Parameters : Scenario 1 }
			\label{scenario_1_tab}
			\begin{tabular}{c c}
				\toprule 
				Parameter & Value\\
				\midrule
				$\Delta T$ & 2 seconds.\\
				$\sigma^1_r$, $\sigma^1_\theta$, $\sigma^1_\phi$ & $100$ m, $2^\circ$, $2^\circ$ \\
				$\sigma^2_r$, $\sigma^2_\theta$, $\sigma^2_\phi$ & $100$ m, $2^\circ$, $2^\circ$ \\
				$\sigma^3_r$, $\sigma^3_\theta$, $\sigma^3_\phi$ & $100$ m, $1.5^\circ$, $1.5^\circ$ \\
				$\tilde{q}_x$  & 0.5 $\text{m}^2/\text{sec}^3$\\
				$\tilde{q}_y$  & 0.5 $\text{m}^2/\text{sec}^3$\\
				$\tilde{q}_z$  & 0.001 $\text{m}^2/\text{sec}^3$\\
				Simulation time & 120 seconds.\\
				\bottomrule
			\end{tabular}	
		\end{table}
		
		The local trackers transmit their estimates every $2$$\Delta T$ to the fusion center, which are then fused together using the existing conservative fusion strategies discussed in the previous sections. The proposed HMD is compared with the AMD and the GMD (covariance intersection) where centralized version of EKF has been chosen as the lower bound. The performance is evaluated on the basis of root mean square error (RMSE) of position and velocities over 500 Monte-Carlo (MC) runs. Consistency of the fusion approaches have also been compared with the help of normalized estimation-error squared (NEES) 
		which is calculated as 
		\begin{align}
			\text{NEES}_k = \frac{1}{M}\sum_{m = 1}^{M}(\hat{\mathbf{x}}^m_{k|k} - \mathbf{x}_k)^T\Gamma^{m^{-1}}(\hat{\mathbf{x}}^m_{k|k} - \mathbf{x}_k).
		\end{align}
		The fused scenario with various tracks are shown in Fig. \ref{fusedScen_1_fig}.
		
\begin{figure}[h!]
			\centering
			\includegraphics[width = \columnwidth]{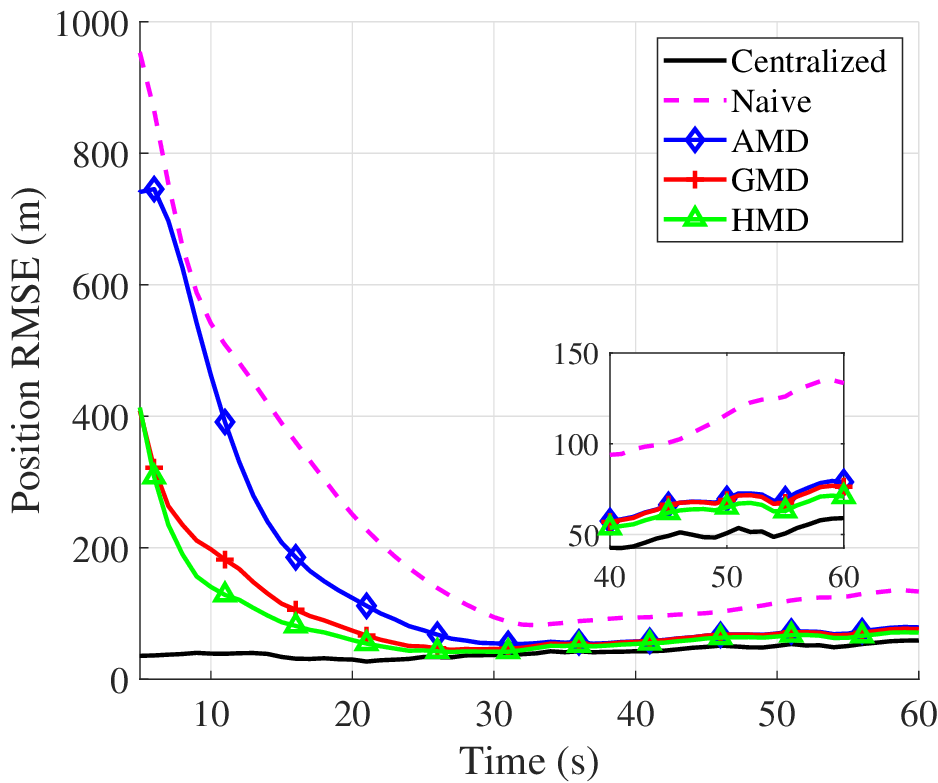}
			\caption{Position root mean square error (RMSE) for scenario 1.}
			\label{rmsePos_1_fig}
		\end{figure}

		The position RMSE plot is shown in the Fig. \ref{rmsePos_1_fig}, where the centralized EKF performs best as expected. It can be seen that the proposed harmonic mean density performs marginally better as compared to the other conservative density approaches. 
		The proposed HMD error plot converges faster than other approaches and suggests an improvement of roughly double when compared to the naive density which starts to diverge at the end of simulation due to rumor propagation. The computation time for Gaussian density fusion for HMD in Fig. \ref{timeGauss_fig} is also almost equal to that of AMD and $50\%$ higher than the covariance intersection.
		
\begin{figure}[h!]
			\centering
			\includegraphics[width = \columnwidth]{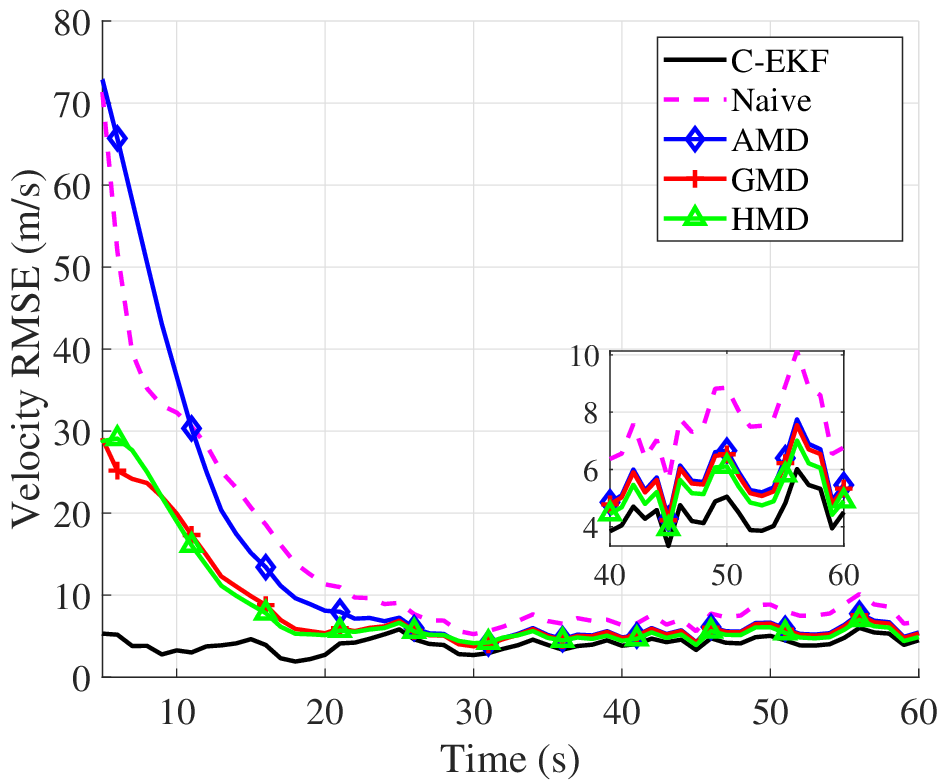}
			\caption{Velocity root mean square error (RMSE) for scenario 1.}
			\label{rmseVel_1_fig}
		\end{figure}

		A similar trend can be seen in the velocity RMSE plot in Fig. \ref{rmseVel_1_fig} with the centralized approach performing best followed by the proposed HMD, GMD, AMD and then the naive fusion. The velocity estimate in case of HMD states an improvement of roughly $50\%$ over naive fusion in terms of accuracy. 
		
\begin{figure}[h!]
			\centering
			\includegraphics[width = \columnwidth]{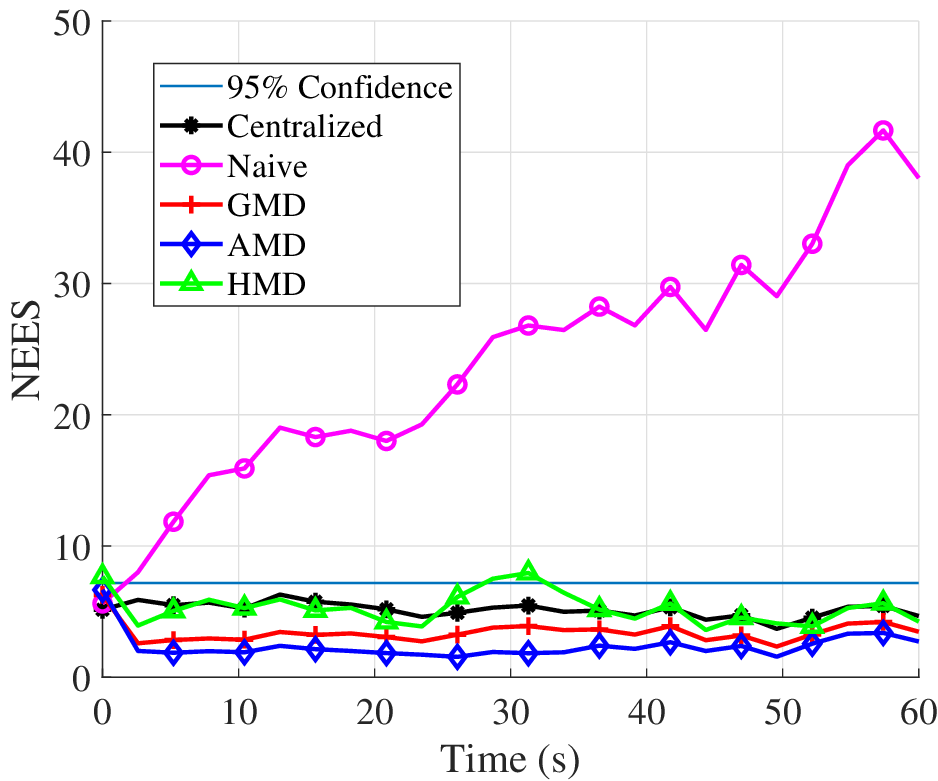}
			\caption{Normalized estimation error squared (NEES) for scenario 1. The HMD can be seen to be fairly consistent along with other strategies.}
			\label{nees_1_fig}
		\end{figure}

		We show the average NEES plotted over 500 MC runs in Fig. \ref{nees_1_fig} to show the consistency of the fused estimates. The $95\%$ confidence interval for $6$ dimension state vector is shown by a blue line in the plot. It can be seen that no point in the naive plot lies inside the confidence region proving its inconsistency. Ignoring the initial transient region, only one point out of 120 time steps lies outside the region in case of HMD, which is acceptable. The other approaches are well below the threshold especially the arithmetic mean density exhibiting its over-conservativeness.  
		
		\subsection{Two Sensor Passive Maneuvering Target Tracking}
\begin{figure}[h!]
			\includegraphics[width = \columnwidth]{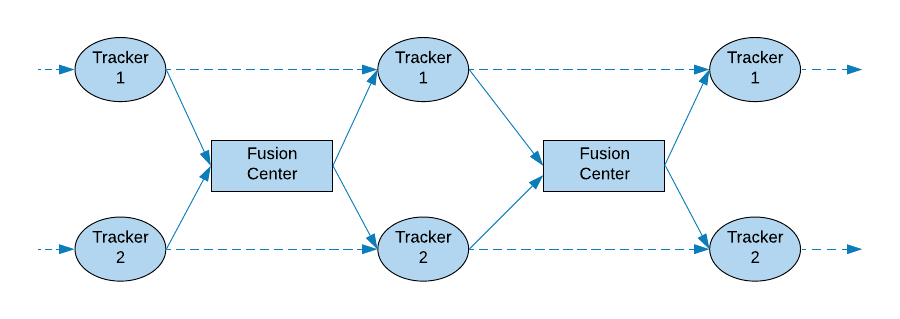}
			\caption{Information flow between fusion center and local trackers for scenario 2.}
			\label{infoFlow_2_fig}
		\end{figure}

		A major advantage of using HMD is the fusion of multi-modal and uni-modal Gaussian densities without using additional machinery like in GMD wherein a separate mechanism is required to compute accurate power of a Gaussian mixture, or AMD where variance deflation techniques are necessary to prevent the estimates from being over-conservative. To illustrate this advantage, a two-sensor multiple model based maneuvering target tracking using bearing-only sensor in 2D was designed. Each local tracker is an IMM consisting of a Gaussian noise NCV model (as discussed in the previous subsection) and a Wiener noise nearly constant acceleration (NCA) model. This scenario is considered non-trivial because of the following reasons :
		
		\begin{itemize}
			\item Unobservability of local trackers.
			\item Different state dimension of the two models used in local IMM tracker (NCA vs NCV).
			\item Feedback requirement from fusion center to improve local estimate.
			\item Motion model mismatch between actual target and the one employed by local trackers.
		\end{itemize}
		To solve the problem of growing fused mixture size, a pruning operation is performed after fusion so that the final mixture consists of two modes. This mixture is sent to the local nodes after each fusion instant as feedback (see Fig. \ref{infoFlow_2_fig}).
		For instance, GMD fusion of two bi-modal Gaussian mixture outputs a fused density with four modes. Since the local IMM trackers employ bi-modal densities, 
		the fused mixture size has to be pruned before transmitting it back. The different dimensions of the two modes is taken care of by padding extra zeros in the estimate of NCV model. This poses problems while using sigma-point Chernoff fusion \cite{gunay2016chernoff} where Cholesky decomposition needs to be carried out, thus, PCF from equation \eqref{pcf} \cite{julier2006empirical} was used instead which is also the fastest method. 
		
\begin{figure}[h!]
			\includegraphics[width = 0.95\columnwidth]{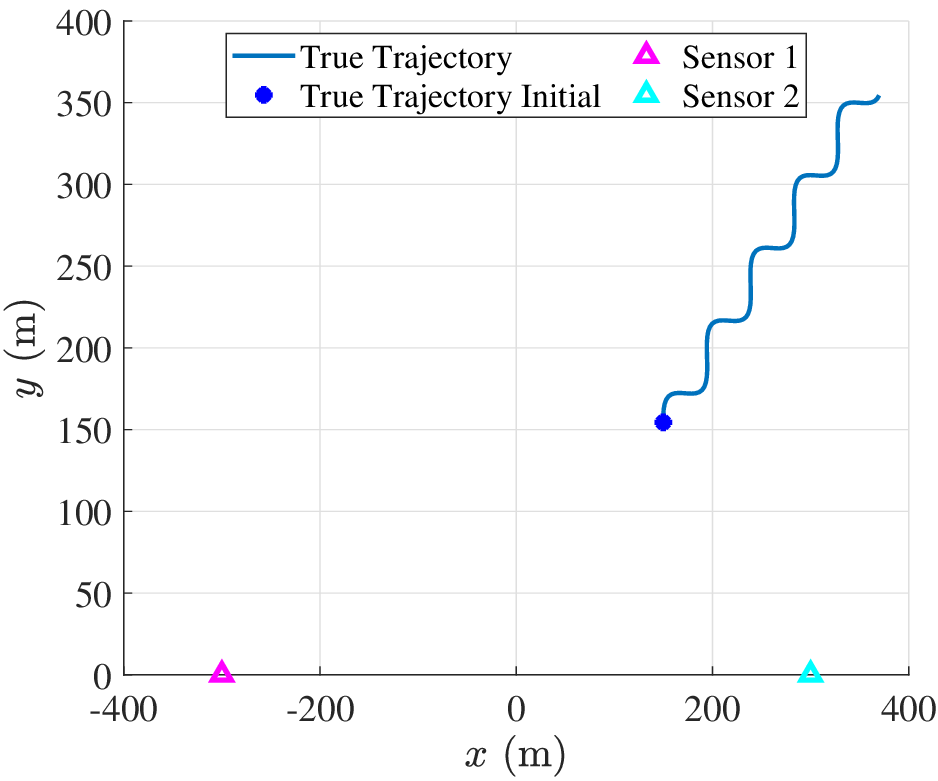}
			\caption{Target and sensor engagements for scenario 2. Note that the sensors are bearings-only.}
			\label{scenario_2_fig}
		\end{figure}

		The scenario is shown in Fig. \ref{scenario_2_fig}. The target moves in a wave like motion about a straight line. Such trajectories are common in underwater warfare engagement scenario and buoyancy-driven underwater gliders \cite{kan2008matlab}. In this case, the target was generated by rotating a sine wave motion in $x$ and $y$ coordinates at a fixed angle. No noise has been added to the true target trajectory due to presence of high maneuvers already. An attempt to track this target using conservative fusion strategies with IMM was made. In addition to the baseline methods mentioned in the previous sections, the centralized methods, centralized-CV and centralized-CA were added to check if non-IMM centralized approaches are suited for the problem. 
		
		The motion model used in the IMM trackers remains the same as mentioned in equation \eqref{modelScen_1_eq} modified to 2D, and with an addition of a separate NCA model categorized by the following parameters,
		\begin{subequations}\label{model_ca_2_eq}
			\begin{align}
				\mathbf{F}_{NCA} &= \begin{bmatrix}
					\mathbf{I}_2 & \Delta T\mathbf{I}_2  & \frac{\Delta T^2}{2}\mathbf{I}_2 \\
					\mathbf{0}_2 & \mathbf{I}_2  & \Delta T \mathbf{I}_2  \\
					\mathbf{0}_2 & \mathbf{0}_2 & \mathbf{I}_2
				\end{bmatrix},\\
				\mathbf{Q}_{NCA}	&= 	\tilde{q}_{NCA}\begin{bmatrix}
					\frac{\Delta T^5}{20}\mathbf{I}_2 & \frac{\Delta T^4}{8}\mathbf{I}_2 & \frac{\Delta T^3}{6}\mathbf{I}_2 \\
					\frac{\Delta T^4}{8}\mathbf{I}_2 & \frac{\Delta T^3}{3}\mathbf{I}_2 & \frac{\Delta T^2}{2}\mathbf{I}_2 \\
					\frac{\Delta T^3}{6}\mathbf{I}_2 & \frac{\Delta T^2}{2}\mathbf{I}_2 & \Delta T \mathbf{I}_2
				\end{bmatrix}.				
			\end{align}
		\end{subequations}
		The sensor measurements are now bearings measured from the true north, thus
		\begin{subequations}
			\begin{align}
				\mathbf{\theta}_k &= \mathbf{h}(x_k,y_k) + \mathbf{v}_k,
			\end{align}
			where,
			\begin{align}
				\mathbf{h}(x_k,y_k) & = \tan^{-1}\left( \frac{x_k}{y_k}\right).
			\end{align}		
		\end{subequations}
		The parameters used in the models are tabulated in Table \ref{scenario_2_tab}. It is to be noted that the target in Fig. \ref{scenario_2_fig} does not adhere  to either NCA or NCV motion model, the plotted trajectory is completely independent of motion statistic used such as the case in real-time target tracking. Therefore, there is a high certainty of errors due to model mismatch. The sensors are located $600$ meters apart and the target starts at $[150\text{m}, 150\text{m}]$ and moves north-east with a velocity of $16$ knots. 
		
\begin{figure}[h!]
			\centering
			\includegraphics[width = \columnwidth]{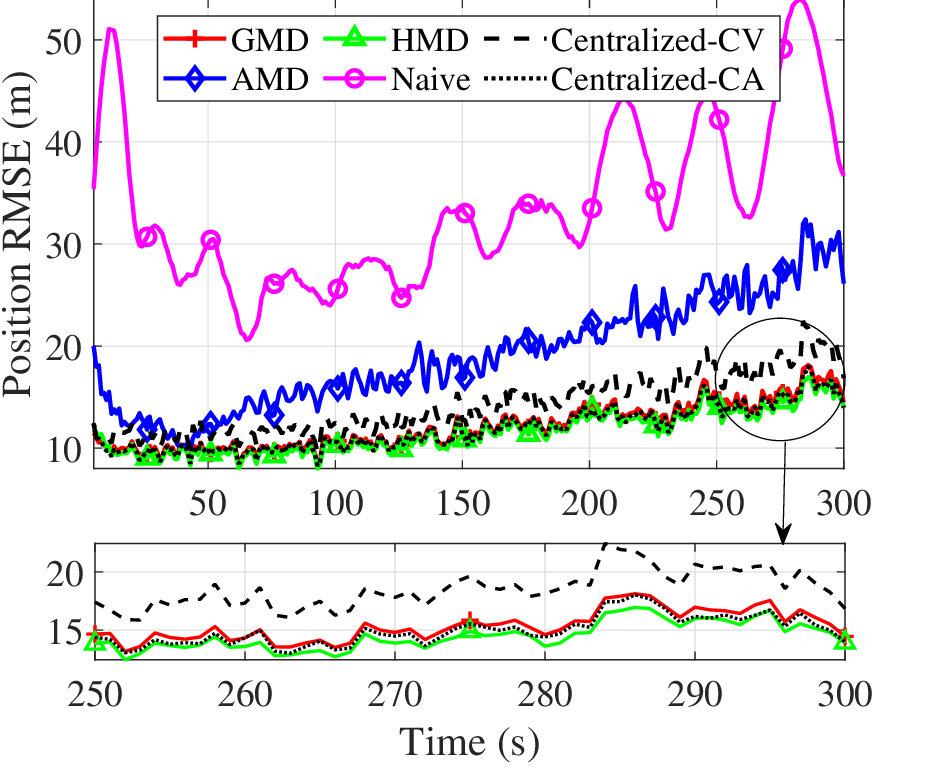}
			\caption{Position root mean square error (RMSE) for scenario 2.}
			\label{rmsePos_2_fig}
		\end{figure}

		The results over 500 Monte-Carlo runs are plotted in Figs. [\ref{rmsePos_2_fig} - \ref{nees_2_fig}]. As expected, the centralized EKF employing NCA model has the least RMSE followed by the proposed HMD. An interesting point to note is that centralized EKF-CV performs poor in terms of RMSE 
		when compared to HMD and GMD, whereas AMD failed to track the target owing to extremely high track-loss\footnote{Track-loss is defined as the divergence of the estimate and is quantified when the final position error $e_f$ exceeds a set threshold $\tau$ (500m in this case),
			\begin{align} e_f = \sqrt{(x_f - \hat{x}_f)^2 + (y_f - \hat{y}_f)^2} \geq \tau,\notag \end{align} where $f$ is final time index. The number of track-loss are usually represented as \% of Monte-Carlo runs. Note that while calculating the RMSE, the diverged tracks are removed} ($> 90\%$) while no track-loss was observed in other techniques.
		\subsubsection*{{Failure of AMD in unobservable systems}}
		In the case of expected a posteriori (EAP) estimation, AMD is an amalgamation which only works when the estimates to be fused are sufficiently close. But since the local nodes in this scenario are non-observable, the \textit{closeness} is not guaranteed. For instance, suppose the two estimates to be fused are $x_1 \sim \mathcal{N}(50, 10)$ and $x_2 \sim \mathcal{N}(-30, 20)$. The resulting densities are  -- $\mathcal{N}(10, 1615)$, $\mathcal{N}(23.39, 6.69)$, $\mathcal{N}(23.33, 13.33)$ and $\mathcal{N}(23.33, 6.67)$  for AMD, HMD, GMD and naive fusion respectively, which shows that the uncertainty in AMD fusion is very high if the proximity between individual means is large. The global observability in this scenario is highly dependent on the quality of feedback since the local nodes are static. Due to a highly uncertain feedback, the quality of local estimate does not improve over time and the fused estimate fails to converge. Amalgamation for unobservable sensors and maneuvering targets is hence avoided due to over conservativeness of the resulting density, inflation of variance is observed. A better strategy would be to use the MAP estimate.
		
\begin{figure}[h!]
			\centering
			\includegraphics[width = \columnwidth]{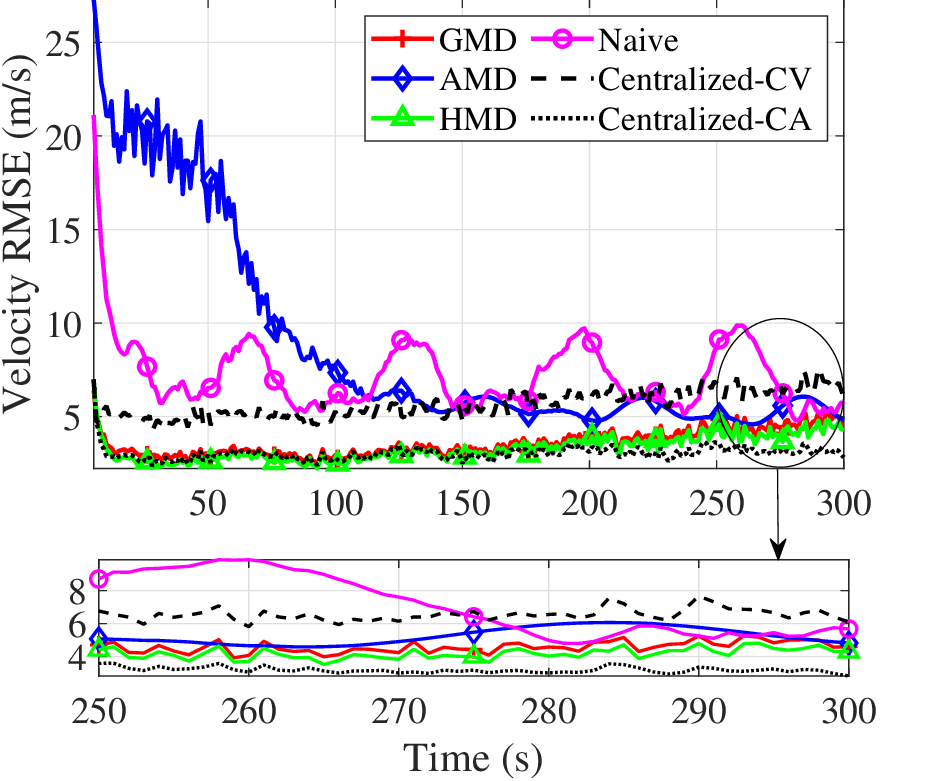}
			\caption{Velocity root mean square error (RMSE) for scenario 2.}
			\label{rmseVel_2_fig}
		\end{figure}

		A similar trend is {\color{black}shown} in the velocity RMSE plot in Fig. \ref{rmseVel_2_fig} where the AMD takes around $75$ time steps to settle down. Harmonic mean density again proves to be the best technique among the mentioned conservative fusion strategies with lowest RMSE in both position and velocity. This is due to a linear approximation of \textit{`common information'} which produces a tighter density with a low variance. When compared to the naive fusion strategy, the average improvement is roughly 2.5 times in position and around 1.5 times in velocity. 
		
\begin{figure}[h!]
			\centering
			\includegraphics[width = \columnwidth]{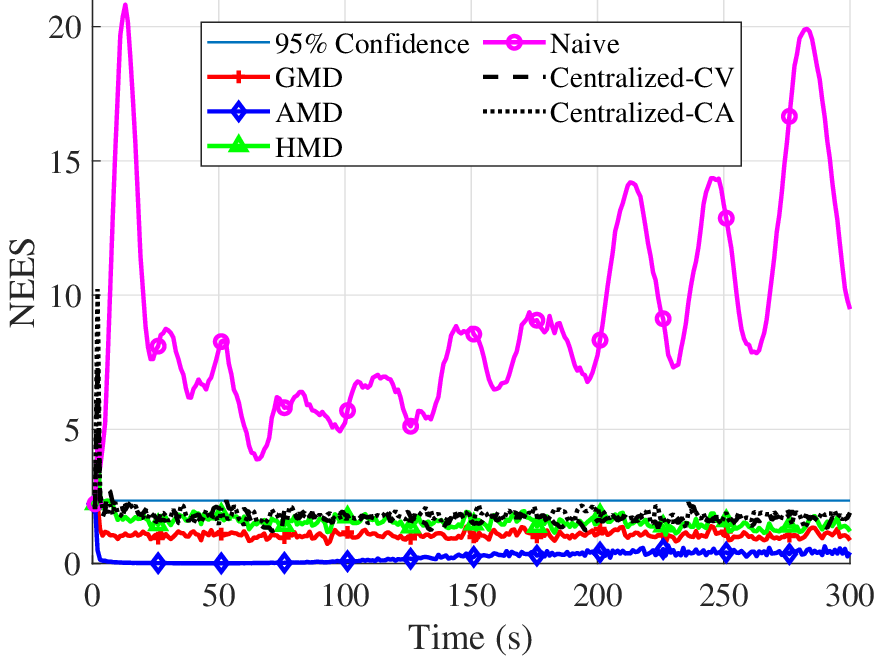}
			\caption{Normalized estimation squared (NEES) - scenario 2.}
			\label{nees_2_fig}
		\end{figure}

		{\color{black}For consistency}, the average NEES over 500 Monte-Carlo runs is plotted in Fig. \ref{nees_2_fig}. Since the fused densities are Gaussian mixtures, the average NEES are calculated using the Gaussian approximation. One-sided 95\% probability is taken as threshold which is shown by a thin blue line in the plot. It can be seen that apart from one point in the initial transient region, there is no marker outside the threshold for 300 time steps which is quite acceptable. {\color{black} Also, the NEES for proposed HMD implementation is closest to that of centralized filters implemented. Note that the central filters are globally-optimal (neglecting non-linear measurements), which prove the effectiveness of the proposed fusion method}. The naive fusion results are totally inconsistent as all points are outside the confidence region. Again, the arithmetic averaging proves to be over conservative with the least average NEES. 
		
\begin{figure}[h!] 
			\begin{subfigure}{0.5\columnwidth}
				\centering
				\includegraphics[width=\columnwidth]{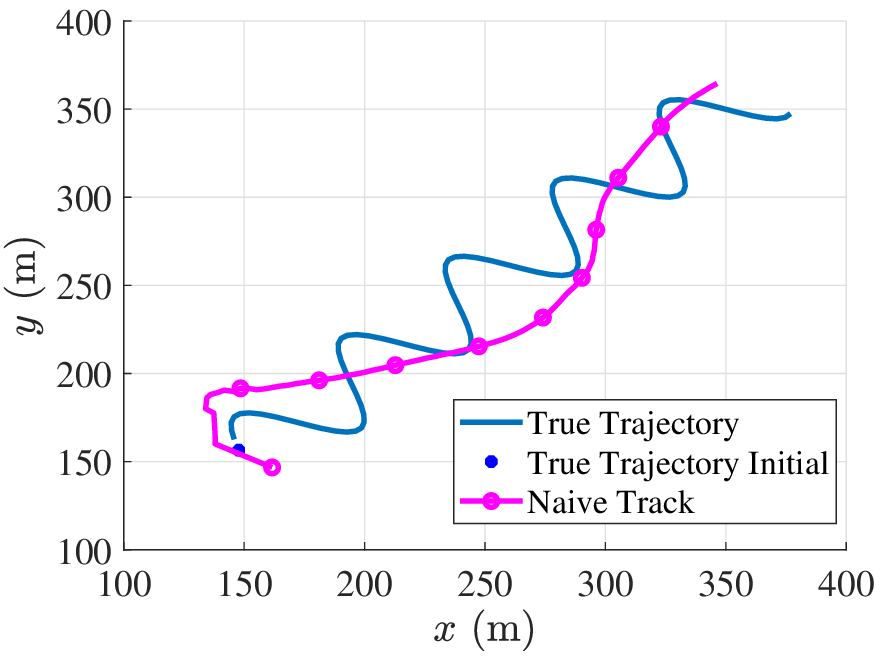}
				\caption{Naive ($\tilde{q}_{NCV} = 0.01$)}
				\label{fig_tracks_2_bot_naive_1}
			\end{subfigure}%
			\begin{subfigure}{0.5\columnwidth}
				\centering
				\includegraphics[width=\columnwidth]{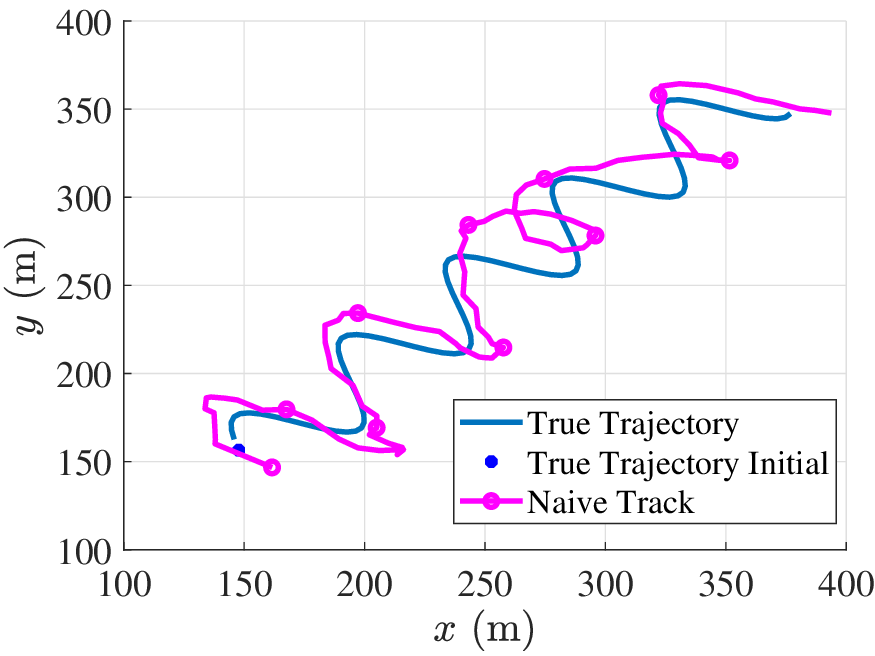}
				\caption{Naive ($\tilde{q}_{NCV} = 0.5$)}
				\label{fig_tracks_2_bot_naive_2}
			\end{subfigure}\\
			\begin{subfigure}{0.5\columnwidth}
				\centering
				\includegraphics[width=\columnwidth]{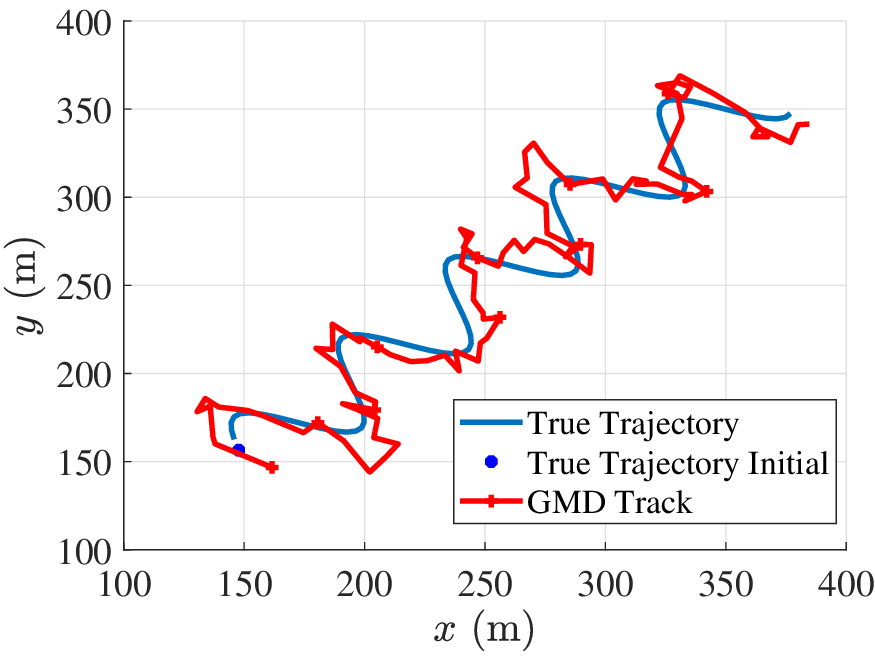}
				\caption{GMD ($\tilde{q}_{NCV} = 0.01$)}
				\label{fig_tracks_2_bot_gmd}
			\end{subfigure}%
			\begin{subfigure}{0.5\columnwidth}
				\centering
				\includegraphics[width=\columnwidth]{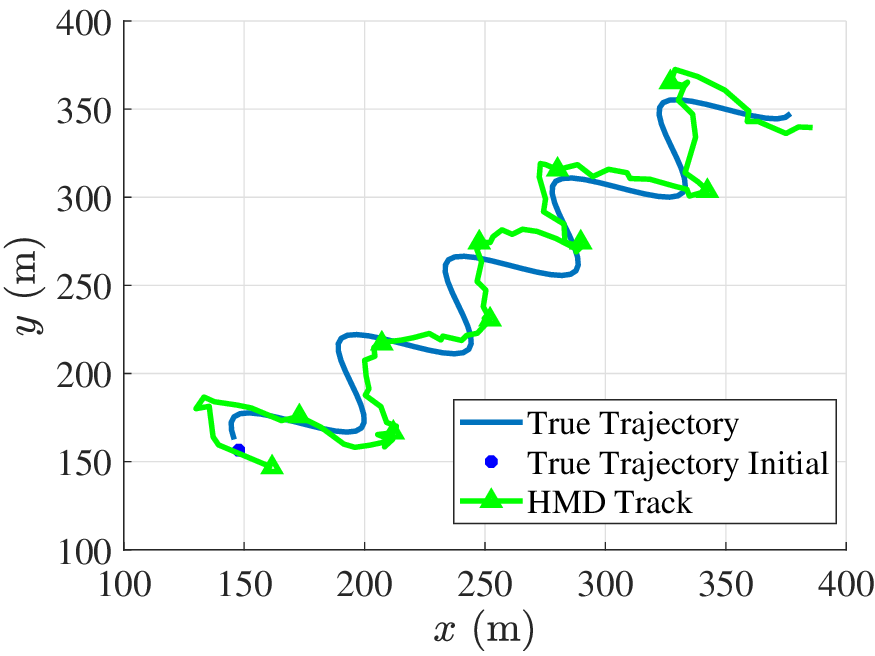}
				\caption{HMD ($\tilde{q}_{NCV} = 0.01$)}
				\label{fig_tracks_2_bot_hmd}
			\end{subfigure}
			\caption{Fused tracks with various approaches for scenario 2. The arithmetic mean density has been excluded due to very high track-loss.}
			\label{fig_tracks_2_bot}
		\end{figure}

		The resulting tracks are shown in Fig. \ref{fig_tracks_2_bot} where the AMD track has been omitted due to high track-loss. Since the naive density results in a lower non-consistent covariance, increasing the process noise improves track quality since it has a similar effect as using a fudge factor \cite{bar2004estimation}. The difference is evident in Fig.  \ref{fig_tracks_2_bot_naive_1} and \ref{fig_tracks_2_bot_naive_2}. The GMD track (Fig. \ref{fig_tracks_2_bot_gmd}) and the HMD track (Fig. \ref{fig_tracks_2_bot_hmd}) faithfully track the target at a lower process noise. 
		
		\begin{table}[h!]
			\centering
			\caption{Parameters : Scenario 2 }
			\label{scenario_2_tab}
			\begin{tabular}{c c}
				\toprule 
				Parameter & Value\\
				\midrule
				$\Delta T$ & 1 second \\
				$R_1$ & $(1.5^\circ)^2$\\
				$R_2$ & $(2^\circ)^2$ \\
				$\tilde{q}_{NCV}$ & $10^{-2}$ $\text{m}^2/\text{sec}^3$ \\
				$\tilde{q}_{NCA}$ & $10^{-3}$ $\text{m}^2/\text{sec}^5$\\
				IMM transition matrix & $\begin{bmatrix} 0.8 & 0.2 \\ 0.8 & 0.2  \end{bmatrix}$\\
				\bottomrule
			\end{tabular}	
		\end{table}
		
		\section{Conclusion}\label{conclusion}
		In this paper, a new strategy that could fuse both uni-modal and multi-modal local track densities without any additions to the framework has been presented. The proposed strategy has been proved to be effective in non-independent fusion architectures, both theoretically and in simulations. We investigated the mathematical properties and derived some useful equations required in the development of a viable fusion method. It was proved that the proposed harmonic mean density does not double count information and an approximate implementation is also presented. Analysis show that HMD is almost as computationally effective as other conservative strategies existing in the literature. 
		The efficacy of the harmonic mean density is presented in two real-life scenarios and shown that the proposed method has the least RMSE among discussed conservative fusion strategies. The NEES results show that the HMD is consistent and is not over-conservative like the arithmetic mean density. A few existing research areas discuss the evaluation of fusion weights and perhaps an accurate implementation without using Gaussian approximation for denominator as in equation \eqref{eqn_Gauss_approx_den}. Nevertheless, the HMD is a promising candidate for fusion of correlated estimates in distributed/decentralized network architectures.

		\appendix
		\section{Proof that the division of Gaussian densities performed in equation \eqref{GaussDivision} is always valid}
		\label{appendix_proof_division}
		The numerator, which is a valid product density has the covariance,
		\begin{subequations}\label{numerator}
			\begin{align}
				\Gamma^{num} &= \left(\Gamma^{i^{-1}}_k + \Gamma_k^{j^{-1}}\right)^{-1}, \notag\\
				&= \Gamma^{i}_k - \Gamma^{i}_k\left( \Gamma^{i}_k + \Gamma^{j}_k\right)^{-1}\Gamma^{i}_k, \\
				& = \Gamma^{j}_k - \Gamma^{j}_k\left( \Gamma^{i}_k + \Gamma^{j}_k\right)^{-1}\Gamma^{j	}_k,
			\end{align}
		\end{subequations}
		where the matrix inversion lemma \cite{bar2004estimation} has been used. Assuming $\Gamma^{i}_k$ and $\Gamma^{j}_k$ are invertible, equation \eqref{numerator} implies that in the sense of positive definiteness, $\Gamma^{num} \preceq \{ \Gamma^{i}_k, \Gamma^{j}_k\}$. Moreover, for the division to be valid, it is required that \cite{acar2020decorrelation},
		\begin{align}
			\Gamma_k^{eq} \succeq \Gamma^{num}_k, \qquad \forall k
		\end{align}
		Using the property of arithmetic average,
		\begin{align}
			\Gamma_k^{eq} \succ \omega\Gamma^i_k + (1-\omega)\Gamma^j_k \succeq \min\{\Gamma^i_k, \Gamma^j_k\}, \label{Corollary_1}
		\end{align}
		we reach $\Gamma_k^{eq} - \Gamma^{num}_k \succ\mathbf{0}, $ $\forall k$ which concludes the proof.

	\bibliographystyle{elsarticle-num}
	\bibliography{main_journal_1.bib}

\end{document}